\documentclass[altaffilletter,aps,nofootinbib,twocolumn,prd,eqsecnum,preprintnumbers,superscriptaddress,10pt,floatfix]{revtex4-2}
\pdfoutput=1
\usepackage{amsmath}
\usepackage{amssymb}
\usepackage{bm}
\usepackage{subfigure}
\usepackage{graphicx}
\usepackage{dcolumn}
\usepackage{booktabs}
\usepackage{amsfonts}
\usepackage{mathtools}
\usepackage{enumerate}
\usepackage{orcidlink}
\usepackage{xcolor}
\usepackage{mathrsfs}
\usepackage{epstopdf}
\usepackage{url}
\usepackage{footnote}
\usepackage{textcomp}
\usepackage{dsfont}
\usepackage{ulem}
\usepackage{enumerate}   
\usepackage{comment}
\usepackage{appendix}
\usepackage{textcomp}
\usepackage{tipa}
\hypersetup{
	colorlinks=true,
	linkcolor=red!70!black,     % \eqref{}, \ref{}, internal links
	citecolor=cyan!99!black,    % \cite{}
	urlcolor=blue!55!black      % URLs
}

\newcommand{\ri}{r_{\mathrm{in}}}\newcommand{\rt}{r_t}\newcommand{\rh}{r_h}
\newcommand{\Mb}{M_{\bullet}}\newcommand{\Ms}{M_{\mathrm{Sch}}}\newcommand{\Mt}{M_{\mathrm{tot}}}\newcommand{\Me}{M_{\mathrm{ext}}}
\newcommand{\NN}{\mathcal N_{\mathrm{sp}}}\newcommand{\gs}{\gamma_{\mathrm{sp}}}

\makeatletter
\newcommand*{\rom}[1]{\expandafter\@slowromancap\romannumeral #1@}
\makeatother

\begin{document}

\title{
A Smoothly Truncated Dark Matter Spike\\around a Black Hole in an Einstein-Cluster Model
}

\author{Hassan Hassanabadi\,\orcidlink{0000-0001-7487-6898}
}
\email{hassanhassanabadi@mail.fresnostate.edu}
\affiliation{Physics Department, California State University, Fresno, CA 93740, USA}

\author{Soroush Zare\,\orcidlink{0000-0003-0748-3386}
}
\email{soroush.z.zare@helsinki.fi}
\affiliation{ Helsinki Institute of Physics, University of Helsinki, P.O. Box 64, FI-00014 Helsinki, Finland}

\begin{abstract}
We introduce a finite dark matter spike around a static, spherically symmetric black hole, with a general power-law index and an Einstein-cluster stress tensor. Smooth inner depletion and outer truncation ensure that the density and its first radial derivative vanish at both boundaries, while a single normalization preserves the prescribed exterior mass. The resulting spacetime contains an inner vacuum gap, a matter region with vanishing radial pressure, and an outer Schwarzschild vacuum. We derive the enclosed mass, density maximum, local logarithmic slope, and matched metric without surface layers, including exact polynomial-root representations of the matter-region metric for spike indices $7/3$ and $3/2$. Energy conditions and radial stability of the constituent circular orbits constrain the admissible configurations. A sufficient bound guarantees these properties for an inner edge at or beyond the Schwarzschild marginal-stability radius and also ensures positivity of the massless-scalar potential. Comparisons at a common mass budget examine how boundary regularity and radial concentration jointly affect the gravitational response. Shadows and light deflection also depend on the global redshift and mass calibration. Scalar waves matched across the three regions with one infinity-normalized time coordinate distinguish local gap-redshift effects from propagation through the complete geometry. The smoothly truncated profile provides a controlled equilibrium model for investigating how finite support and mass redistribution affect black hole observables; its taper scales, formation, and collective stability require additional physical input.
\end{abstract}

\maketitle

%%%%%%%%%%%%%%%
%%%%%%%%%%%%%%%
%%%%%%%%%%%%%%%

\section{Introduction}\label{sec.intro}
The distribution of dark matter close to a black hole connects galactic structure with strong-field gravity. Large-scale halo models constrain the surrounding mass distribution, but do not uniquely determine its continuation toward the horizon~\cite{Bertone2005,Salucci2019}. In particular, a power-law enhancement near the center must be distinguished from the global halo in which it is embedded. For a relativistic model, specifying this enhancement also raises questions about its inner support, finite mass, stress tensor, and matching to the surrounding spacetime.

Black hole growth can produce a steep central spike, but its survival depends on the history of the system. The adiabatic construction provides a reference power law, whereas different seed histories can weaken the enhancement~\cite{Gondolo1999,Ullio2001}. Mergers, stellar heating, capture, and subsequent evolution can substantially redistribute the central dark matter~\cite{Merritt2002,Merritt2004,Bertone2005b}. Dark matter spikes have also been studied as environments for primordial-black-hole mergers. In  ~\cite{Nishikawa2019} estimated the merger rate for spikes embedded in galactic halos and showed that it can be strongly reduced by relaxation when replenishment is inefficient. Consequently, a prescribed equilibrium density is useful for isolating gravitational effects, but its slope and cutoff should not be identified with a unique formation outcome. This distinction is particularly important when introducing suppression factors whose mathematical properties are imposed rather than derived dynamically.

The innermost profile requires special care because relativistic capture modifies the available particle orbits. Relativistic distribution-function calculations incorporate capture and, in rotating backgrounds, its dependence on spin~\cite{Sadeghian2013,Ferrer2017}; annihilation evolution can instead produce a weak inner cusp~\cite{Shapiro2016}. These results show that a pure power law is not generally maintained down to an arbitrarily chosen inner radius. The behavior at the inner edge must therefore be related to the orbital population and the processes that remove or redistribute particles. A self-consistent spacetime additionally requires the stresses associated with that population, rather than the density alone~\cite{Xiong2026}.

Einstein clusters provide a specific equilibrium description in which circular particle orbits generate tangential support with vanishing radial pressure, originating in Einstein's spherical circular-orbit construction~\cite{Einstein1939}. This framework has been applied to galactic halos and black hole environments with different density prescriptions~\cite{Bohmer2007,Jusufi2023,Acharyya2024}. Relativistic black hole--halo constructions and energy-condition analyses constrain which configurations are admissible~\cite{Shen2024,Shen2025}. In spiky Einstein clusters, the mass onset at the inner edge affects the location of the marginally stable support boundary~\cite{Maeda2025}. It is therefore necessary to distinguish the prescribed density edge, the existence of timelike constituents, and their radial stability. Even when individual circular orbits are stable in the fixed geometry, collective stability under matter and metric perturbations remains a separate question~\cite{Maeda2025,Geralico2012}.

The motivation for controlling these ingredients is observational as well as geometrical. Stellar-orbit studies constrain extended mass near the Galactic-center black hole~\cite{Lacroix2018,Shen2024b,GRAVITY2024}, while optical calculations relate the density and metric to shadow properties~\cite{Nampalliwar2021,Daghigh2024}. Perturbation studies further show that ringdown and gravitational-wave propagation depend on the surrounding geometry and matter description~\cite{Daghigh2022,Zhao2023,Speeney2024,Zhao2024}. Inspiral calculations provide another probe of dense environments, although dynamical feedback can change the initially assumed spike~\cite{Eda2013,Eda2015,Kavanagh2020}. These applications motivate a self-consistent background that connects the assumed density and stresses to the gravitational response.

The depleted-halo construction of Hassanabadi and Zare~\cite{Hassanabadi2026} assigns the reference mass inside a prescribed inner edge to the central black hole, leaving a vacuum gap between its horizon and the surviving halo. We adopt this depletion and mass-allocation prescription for a finite spike, with the exterior density and outer boundary specified below.

The contribution of this work is a finite, smoothly tapered power-law spike that combines a prescribed exterior mass budget, an explicitly matched three-region Einstein-cluster geometry, and analytic admissibility conditions. Squared inner-depletion and outer-truncation factors set the boundary behavior, while a single normalization preserves the chosen exterior reference mass. This construction connects the profile shape and boundary behavior to a fixed mass budget. Its density and first derivative vanish at both vacuum edges, while Einstein's equations determine the accumulated mass, metric, and tangential pressure. We check the spacetime junction conditions separately from density regularity, since a finite density jump is not equivalent to a thin shell. The resulting model describes a static equilibrium with specified support, rather than a dynamical calculation of capture or tidal truncation.

We use this background to examine energy conditions, circular-orbit stability, shadows, regional light deflection, photon-orbit wave diagnostics, third-order scalar WKB estimates, and matched real-frequency scattering. A matched comparison with the Einstein-cluster examples of Ref.~\cite{Maeda2025} holds the central mass, exterior matter mass, and outer radius fixed. The comparison examines the combined effects of the radial distribution and boundary prescription at a common mass budget; it does not independently isolate smoothness from concentration or global redshift.

The paper is organized as follows.  
Section~\ref{sec:density} introduces the density, mass allocation, and normalization. Section~\ref{sec:shape} establishes the density maximum, power-law interval, and boundary regularity. Sections~\ref{sec:metric} and~\ref{sec:admissibility} derive the regional spacetime, matching, and physical admissibility conditions. Section~\ref{sec:geodesics} discusses circular orbits, shadows, and matched profile comparisons. Sections~\ref{sec:lensing} and~\ref{sec:waves} examine light deflection and scalar-wave diagnostics, respectively. The conclusion summarizes the model's implications and limitations. Throughout this work, we use geometrized units $G=c=1$ and the metric signature $(-,+,+,+)$.

%%%%%%%%%%%%%%%
%%%%%%%%%%%%%%%
%%%%%%%%%%%%%%%

\section{Spike density, mass budget, and normalization}\label{sec:density}

Motivated by the power-law description of black hole-induced dark matter spikes~\cite{Gondolo1999,Ullio2001}, we adopt the reference density
\begin{eqnarray}
	\rho_{\mathrm{ref}}(r)=\rho_s(r_{\mathrm{sp}}/r)^{\gs},
\end{eqnarray}
{where $\gamma_{\rm sp}$ is the spike power-law index, restricted to $0<\gamma_{\rm sp}<3$. The density amplitude is defined by}
\begin{eqnarray}
	\mathcal A_\rho=4\pi\rho_s r_{\mathrm{sp}}^{\gs},%\qquad 0<\gs<3.	
\end{eqnarray}
This quantity is distinct from the dimensionless inner-edge parameter $\kappa$ introduced below. Only the combination $\rho_s r_{\mathrm{sp}}^{\gs}$ sets this amplitude. We set $r_{\mathrm{sp}}=\rt$ as a reference scale; it is not an independently derived influence radius. The inner vacuum gap follows the depleted-halo prescription of Ref.~\cite{Hassanabadi2026}. Here we apply it to a finite power-law spike and impose squared tapers at the inner and outer support edges. The modified density is
\begin{equation}\label{density}
	\rho(r)=\begin{cases}
		0,&\rh<r\leq\ri,\\[1mm]
		\dfrac{\NN\mathcal A_\rho(r-\ri)^2(\rt-r)^2}{4\pi\rt^2r^{\gs+2}},
		&\ri<r<\rt,\\[2mm]
		0,&r\geq\rt,
	\end{cases}
\end{equation}
The prescribed $\rho(r)$ is the energy density entering the stress tensor and Einstein equations below. Here $r_{\mathrm{in}}=\kappa r_h$, with $\kappa>1$, sets the inner edge, and $r_t>r_{\mathrm{in}}$ fixes the outer boundary.
At fixed radius and amplitude, the middle expression has the local untapered limit $\rho(r)\propto r^{-\gs}$ as $\ri\to0$ and $\rt\to\infty$; the finite support intervals are retained in the present model.

An approximately power-law interval emerges when the support boundaries are well separated. The outer radius is a model input, not a truncation scale derived from galactic dynamics.

We retain a general index $0<\gamma_{\mathrm{sp}}<3$, using $\gamma_{\mathrm{sp}}=7/3$, familiar from adiabatic-growth spike models~\cite{Gondolo1999}, as the benchmark. The taper factors redistribute the reference density within the prescribed support. We determine their normalization by requiring the surviving spike to contain the assigned exterior mass.

To fix the mass allocation, prescribe the reference mass $M_H$ inside $\rt$:
\begin{subequations}
	\begin{eqnarray}
		M_H=4\pi\int_0^{\rt}\rho_{\mathrm{ref}}(x)x^2dx=\frac{\mathcal A_\rho\rt^{3-\gs}}{3-\gs},
	\end{eqnarray}
	with
	\begin{eqnarray}
		\mathcal A_\rho=\frac{(3-\gs)M_H}{\rt^{3-\gs}}.
	\end{eqnarray}
\end{subequations}
The amplitude $\mathcal A_\rho$ is the same quantity defined above and is fixed by $M_H$, $r_t$, and $\gamma_{\mathrm{sp}}$. The reference cusp is integrable at the origin for $\gs<3$. It is used for bookkeeping; Eq.~\eqref{density}, not its extrapolation to zero, is the actual matter distribution outside the final hole. No static matter is assumed inside the horizon.

Following the mass-allocation prescription of Ref.~\cite{Hassanabadi2026}, we adopt a maximal-absorption convention: the reference mass initially assigned to radii below the final inner edge is incorporated into the central black hole. This assigned mass is
\begin{equation}\label{eq:absorbed}
	M_{\mathrm{abs}}=\frac{\mathcal A_\rho\ri^{3-\gs}}{3-\gs}
	=M_H(\ri/\rt)^{3-\gs}.
\end{equation}
The central mass is therefore $\Mb=\Ms+M_{\mathrm{abs}}$, with horizon radius $\rh=2\Mb$ and prescribed matter edge $\ri=\kappa\rh$. Since the assigned mass depends on this edge, these relations must be solved consistently. Substitution gives the implicit horizon equation
\begin{equation}\label{horizon}
	\rh=2\Ms+2M_H(\kappa\rh/\rt)^{3-\gs}.
\end{equation}
Choose the physical branch connected continuously to $\rh=2\Ms$ as $M_H\to0$, with $\ri<\rt$. The remaining mass and total gravitational mass are
\begin{eqnarray}\label{massbudget}
	\nonumber
	\Me&=&\frac{\mathcal A_\rho}{3-\gs}(\rt^{3-\gs}-\ri^{3-\gs})\\
	&=&M_H-M_{\mathrm{abs}},\\
	\Mt&=&\Mb+\Me=\Ms+M_H.
\end{eqnarray}
This is an assumed final mass allocation, not an adiabatic-invariant calculation or a time-dependent capture solution. These gravitational masses enter Einstein's equations; they are not particle rest masses integrated over proper volume.

The normalization, coefficient, primitive, and accumulated-mass expressions below are valid for every $\gs$ in the adopted range $0<\gs<3$, including $7/3$. At $\gs=1$ or $2$ a zero-exponent primitive term becomes logarithmic. The exterior integral itself remains finite for any finite real exponent because $\ri>0$, but the reference mass integral from zero used here would diverge at $\gs\geq3$.

Equation~\eqref{massbudget} fixes the exterior mass before tapering. The following integral determines the normalization that preserves that mass after tapering:
\begin{equation}\label{normalizationintegral}
	\Me=\NN\mathcal A_\rho\int_{\ri}^{\rt}x^{2-\gs}
	\left(1-\frac{\ri}{x}\right)^2\left(1-\frac x{\rt}\right)^2dx.
\end{equation}
Let $a=\ri$ only within the following primitive, and introduce
\begin{equation}
	\begin{aligned}
		c_{-2}&=a^2,\qquad c_{-1}=-2a-2a^2/\rt,\\
		c_0&=1+4a/\rt+a^2/\rt^2,\\
		c_1&=-2/\rt-2a/\rt^2,\qquad c_2=1/\rt^2.
	\end{aligned}
\end{equation}
An exact primitive is
\begin{subequations}
	\begin{eqnarray}\label{primitive}
		F_{\gs}(r)&=&\sum_{j=-2}^{2}c_j\frac{r^{3-\gs+j}}{3-\gs+j},\\
		F_{\gs}'(r)&=&r^{2-\gs}(1-a/r)^2(1-r/\rt)^2.
	\end{eqnarray}
\end{subequations}
When a denominator vanishes, replace that term by $c_j\ln(r/r_{\mathrm{ref}})$, where $r_{\mathrm{ref}}>0$ is an arbitrary reference length. Only differences of primitives enter the mass, so an arbitrary logarithmic reference scale cancels.

Consequently,
\begin{equation}\label{normalization}
	\NN=\frac{\rt^{3-\gs}-\ri^{3-\gs}}
	{(3-\gs)[F_{\gs}(\rt)-F_{\gs}(\ri)]}>1.
\end{equation}
The inequality follows because both edge factors lie below one in the occupied interval. At $M_H=0$, the normalization can be assigned its finite continuous limit, while the actual density and exterior mass vanish. The exact vacuum limit is $m(r)=M_{\mathrm{Sch}}$, $A(r)=B(r)=1-2M_{\mathrm{Sch}}/r$, and $C=1$. The accumulated spike mass is
\begin{equation}
	h(r)=\begin{cases}
		0,&\rh\leq r\leq\ri,\\
		\NN\mathcal A_\rho[F_{\gs}(r)-F_{\gs}(\ri)],&\ri<r<\rt,\\
		\Me,&r\geq\rt,
	\end{cases}
\end{equation}
The total enclosed mass is $m(r)=M_\bullet+h(r)$.
It obeys $m'(r)=4\pi r^2\rho(r)\geq0$ and $h(\rt)=\Me$. 

Unless otherwise stated, numerical examples use $M_{\mathrm{Sch}}=1$, $M_H=0.01$, $r_t=10^4$, $\kappa=3$, and $\gamma_{\mathrm{sp}}=7/3$. Lengths and masses are expressed in $M_{\mathrm{Sch}}$ units and energy density in $M_{\mathrm{Sch}}^{-2}$, unless a central-mass normalization is stated. These are illustrative inputs rather than an observational fit; Table~\ref{tab:sensitivity} varies the density index.

The $\gamma_{\mathrm{sp}}=7/3$ row of Table~\ref{tab:sensitivity} lists the resulting central mass and normalization. The horizon and inner radii follow from $r_h=2M_\bullet$ and $r_{\mathrm{in}}=\kappa r_h$, while the exterior and total masses follow from the mass budget above.
The reference amplitude is $\mathcal A_\rho=1.436289793\times10^{-5}$ in these units. Direct integration of the normalized density in logarithmic radius agrees with $M_{\mathrm{ext}}$ to an absolute residual below $10^{-11}$; this checks the mass integral, not the assumed absorption history.

%%%%%%%%%%%%%%%
%%%%%%%%%%%%%%%
%%%%%%%%%%%%%%%

\section{Spike interval, density maximum, and edge regularity}\label{sec:shape}

We identify the approximate power-law interval and unique density maximum, then establish the boundary regularity used in metric matching.

\begin{figure*}[!t]\centering
	\includegraphics[width=\linewidth]{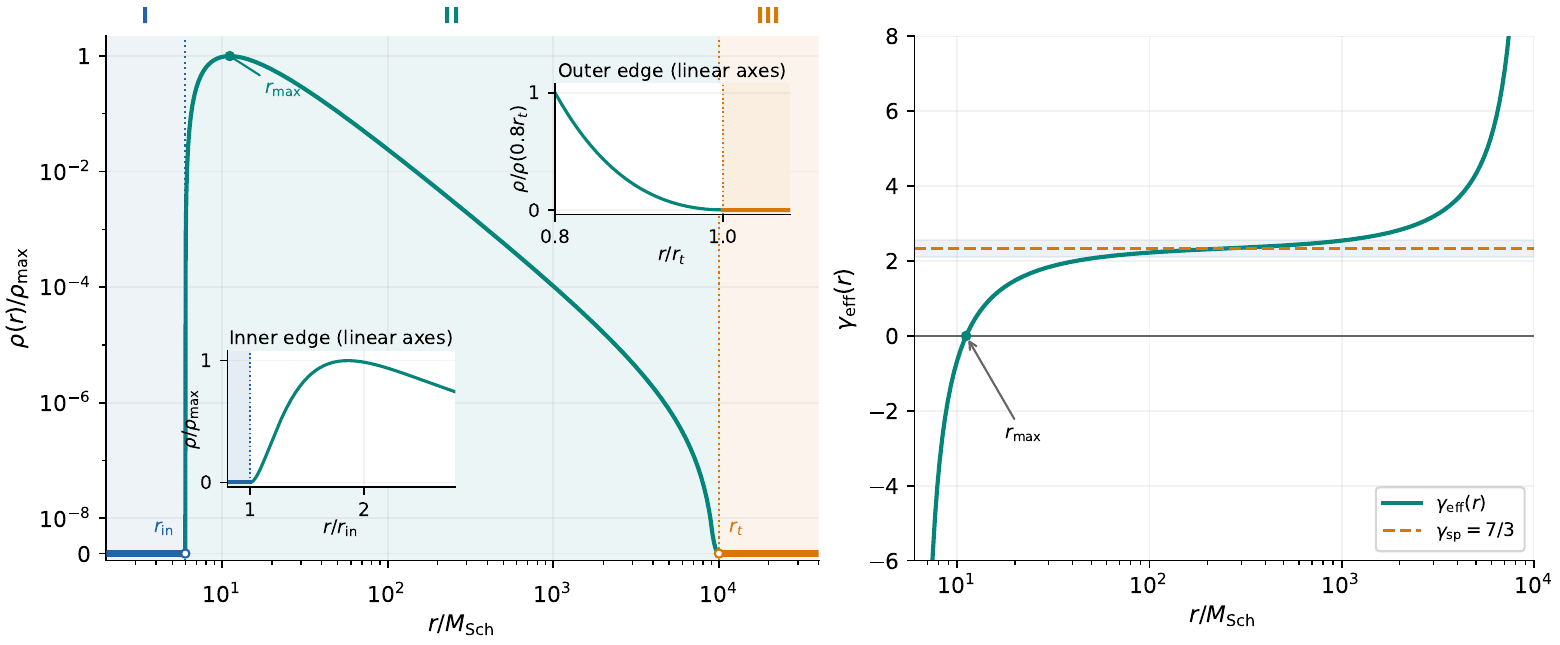}
	\caption{Regional density and effective slope of the benchmark spike. Left: density normalized by its maximum, with blue, green, and orange identifying region I (vacuum gap), region II (spike), and region III (outer vacuum). The vacuum segments lie exactly at zero. A symmetric-logarithmic vertical scale is linear near zero ($\rho/\rho_{\max}\lesssim10^{-8}$) and logarithmic at larger positive densities; no density floor is introduced. The peak and support edges are marked. The insets use linear axes to resolve the quadratic inner and outer tapers and their zero-slope joins to vacuum. Right: $\gamma_{\mathrm{eff}}(r)$ from Eq.~\eqref{eq:slope}, defined only within the occupied interval. The shaded band lies within ten percent of $7/3$; negative values describe rising density before $r_{\max}$. The finite vertical range clips the divergent endpoint slopes.}\label{fig:density}
\end{figure*}

To identify the part of the profile that behaves as a spike, consider its local logarithmic slope,
\begin{equation}\label{eq:slope}
	\gamma_{\mathrm{eff}}(r)=-\frac{d\ln\rho(r)}{d\ln r}
	=\gs-\frac{2\ri}{r-\ri}+\frac{2r}{\rt-r}.
\end{equation}
Thus $\rho(r)\propto r^{-\gs}$ approximately when $\ri\ll r\ll\rt$. The slope equals $\gs$ at $r=\sqrt{\ri\rt}$ by cancellation of the taper terms. Equality at a single radius is insufficient: the benchmark retains slopes within about ten percent of $7/3$ across approximately $60\lesssim r/M_{\mathrm{Sch}}\lesssim1000$. The right panel of Fig.~\ref{fig:density} identifies this interval.

The logarithmic slope also determines the density maximum. Since $\rho(r)>0$ within the support, $\rho'(r)=0$ is equivalent to $\gamma_{\mathrm{eff}}(r)=0$. Moreover,
\begin{equation}
	\frac{d\gamma_{\mathrm{eff}}(r)}{dr}=\frac{2\ri}{(r-\ri)^2}+\frac{2\rt}{(\rt-r)^2}>0.
\end{equation}
The slope increases from $-\infty$ to $+\infty$, so its unique zero determines the density maximum. Solving $\gamma_{\mathrm{eff}}(r_{\max})=0$ gives the physical root in the numerically stable form
\begin{equation}\label{eq:peak}
	r_{\max}=\frac{2(\gs+2)\ri\rt}{\gs(\ri+\rt)+\sqrt{\gs^2(\ri-\rt)^2+16\ri\rt}}.
\end{equation}
This form holds throughout $0<\gs<3$, including $\gs=2$. For $\rt\gg\ri$, it approaches $(1+2/\gs)\ri$, or $13\ri/7$ for $7/3$. For the benchmark, $r_{\max}=11.138738591M_{\mathrm{Sch}}$ and $\rho_{\max}=2.059287299\times10^{-9}M_{\mathrm{Sch}}^{-2}$. Figure~\ref{fig:density} shows the depleted edge, density maximum, and slope across the occupied interval.

For a small positive displacement $\varepsilon$, the density approaches the two vacuum boundaries as
\begin{subequations}
	\begin{eqnarray}\label{eq:edgeexpansion}
		\rho(r_{\mathrm{in}}+\varepsilon)&=&c_{\mathrm{in}}\varepsilon^2+O(\varepsilon^3),\\
		\rho(r_t-\varepsilon)&=&c_t\varepsilon^2+O(\varepsilon^3).
	\end{eqnarray}
\end{subequations}
The coefficients are $c_e=\mathcal N_{\mathrm{sp}}\mathcal A_\rho r_e^{-\gamma_{\mathrm{sp}}-2}(1-r_{\mathrm{in}}/r_t)^2/(4\pi)$, for $e=\mathrm{in},t$, and are positive for a nonzero spike. These expansions
give $\rho(\ri)=\rho'(\ri)=\rho(\rt)=\rho'(\rt)=0$. Thus the density joins vacuum with a continuous first derivative. The second derivative generally jumps. Integrating the density then gives
\begin{subequations}\label{eq:massedges}
	\begin{eqnarray}
		h(\ri)&=&h'(\ri)=h''(\ri)=0,\\
		h(\rt)&=&\Me, \quad h'(\rt)=h''(\rt)=0.
	\end{eqnarray}
\end{subequations}
These relations supply the boundary data used in metric matching. The linear insets in Fig.~\ref{fig:density} resolve the quadratic joins to vacuum.

%%%%%%%%%%%%%%%
%%%%%%%%%%%%%%%
%%%%%%%%%%%%%%%

\section{Einstein equations, regional metric, and boundary matching}\label{sec:metric}

Let $i=I,II,III$ denote the gap, spike, and outer-vacuum domains of Eq.~\eqref{density}. Regional subscripts denote restrictions of one matched spacetime, with common coordinates $t$ and $r$. Einstein's equations, with vanishing cosmological constant, are
\begin{equation}\label{eq:einsteinfull}
	G^{(i)}_{\mu\nu}=8\pi T^{(i)}_{\mu\nu}.
\end{equation}
Let $A_i(r)$ and $B_i(r)$ denote the temporal and radial metric functions in region $i$. The lapse is $\sqrt{A_i(r)}$, and $B_i(r)=1-2m_i(r)/r$ defines the enclosed gravitational mass $m_i(r)$. With the regional coordinates specified above, the metric and stress tensor are
\begin{eqnarray}\label{eq:ansatz}
	ds^2=-A_i(r)dt^2+\frac{dr^2}{B_i(r)}+r^2d\Omega^2,
\end{eqnarray}
\begin{eqnarray}
	T^{(i)\mu}{}_{\nu}(r)=\operatorname{diag}[-\rho_i(r),p_{r,i}(r),p_{t,i}(r),p_{t,i}(r)].
\end{eqnarray}
Here $\rho_i(r)=T^{(i)}_{\mu\nu}(r)u_i^\mu(r)u_i^\nu(r)$ is the energy density measured by static observers, where $u_i^\mu(r)=A_i(r)^{-1/2}(1,0,0,0)$. It is not the particle rest-mass density. The spike has $\rho_{II}(r)=\rho(r)$ from Eq.~\eqref{density}; regions I and III have $\rho_i=p_{r,i}=p_{t,i}=0$. Circular constituents in region II give $p_{r,II}=0$~\cite{Einstein1939,Bohmer2007,Jusufi2023,Maeda2025}, so $p_{r,i}=0$ in all regions. The density and radial-pressure equations take the standard spherical Einstein-cluster form~\cite{Bohmer2007,Maeda2025}:
\begin{subequations}
	\begin{eqnarray}
		8\pi\rho_i(r)&=&\frac{1-B_i(r)}{r^2}-\frac{B_i'(r)}r,\\
		0=8\pi p_{r,i}(r)&=&\frac{B_i(r)-1}{r^2}+\frac{B_i(r)A_i'(r)}{rA_i(r)}.
	\end{eqnarray}
\end{subequations}
Using the enclosed-mass definition above reduces these equations to
\begin{subequations}\label{eq:fieldode}
	\begin{eqnarray}
		m_i'(r)&=&4\pi r^2\rho_i(r),\\
		\frac{A_i'(r)}{A_i(r)}&=&\frac{2m_i(r)}{r[r-2m_i(r)]}.
	\end{eqnarray}
\end{subequations}
In the vacuum domains, $m_I=M_\bullet$ and $m_{III}=M_{\mathrm{tot}}$ are constant; integration gives $B_i(r)=1-2m_i/r$ and $A_i(r)=C_i[1-2m_i/r]$. The common time normalization fixes $C_{III}=1$, while inner matching fixes $C_I=C$. In region II, $m_{II}(r)=M_\bullet+h(r)$ and the same radial equation determines $A_{II}(r)$.

The anisotropic conservation equation~\cite{Bowers1974}, implied by the contracted Bianchi identity, reads
\begin{equation}\label{eq:conservation}
	\begin{aligned}
		p_{r,i}'(r)={}&-\frac{A_i'(r)}{2A_i(r)}[\rho_i(r)+p_{r,i}(r)]\\
		&+\frac2r[p_{t,i}(r)-p_{r,i}(r)].
	\end{aligned}
\end{equation}
With $p_{r,i}=0$, the tangential stress is therefore
\begin{equation}\label{eq:pt}
	p_{t,i}(r)=\frac{\rho_i(r)m_i(r)}{2[r-2m_i(r)]}.
\end{equation}
This vanishes in regions I and III and supplies the tangential support in region II. The angular Einstein equation then follows consistently from the other field equations and conservation; no independent tangential equation of state is imposed.

Integrating Eq.~\eqref{eq:fieldode} with the previously obtained mass gives the exact regional functions. The vacuum--matter--vacuum construction follows the finite-support Einstein-cluster setting of Ref.~\cite{Maeda2025}; the inner depletion and mass allocation follow Ref.~\cite{Hassanabadi2026}. The particular metric below is obtained from our tapered density, with junction conditions checked following Ref.~\cite{Israel1966}.
The global time is normalized by $A(\infty)=1$. The three regional functions are
%\begin{subequations}
\begin{eqnarray}
	A_I(r)&=&C(1-r_h/r),\quad B_I(r)=1-r_h/r,\\
	A_{II}(r)&=&A_H(r),\quad B_{II}(r)=1-2m_{II}(r)/r,\\
	A_{III}(r)&=&B_{III}(r)=1-2M_{\mathrm{tot}}/r.
\end{eqnarray}
%\end{subequations}
Here $C$ fixes the inner redshift relative to infinity. Integrating the radial field equation from infinity and evaluating the vacuum contributions determines this factor from the matter-region integral and the boundary values of $B_I(r)$ and $B_{III}(r)$:
\begin{equation}\label{eq:gapNormalization}
	C=\frac{B_{III}(r_t)}{B_I(r_{\mathrm{in}})}
	\exp\left[-\int_{r_{\mathrm{in}}}^{r_t}
	\frac{2m_{II}(x)}{x[x-2m_{II}(x)]}\,dx\right].
\end{equation}
Thus $C$ is fixed by the mass distribution and $A(\infty)=1$, not an independent parameter. Region III has no additional multiplier because the outer time normalization already fixes it.

The exact matter-region metric function is
\begin{equation}\label{Ahalo}
	A_H(r)=\left(1-\frac{2\Mt}{\rt}\right)
	\exp\left[-\int_r^{\rt}\frac{2m(x)}{x[x-2m(x)]}\,dx\right].
\end{equation}
At $r=r_{\mathrm{in}}$, Eq.~\eqref{Ahalo} and Eq.~\eqref{eq:gapNormalization} give $A_H(r_{\mathrm{in}})=C B_I(r_{\mathrm{in}})$, ensuring the inner boundary value agrees with the vacuum gap.

The matter-region integral also admits elementary primitives after rationalizing the radius for selected indices. We give the benchmark $\gamma_{\mathrm{sp}}=7/3$ and the simpler alternative $\gamma_{\mathrm{sp}}=3/2$, already included in Table~\ref{tab:sensitivity}. These exact representations involve polynomial roots rather than radial quadrature; the roots generally require numerical evaluation.

For either index, introduce $\eta=r_{\mathrm{in}}/r_t$, $y=(r/r_t)^{1/s}$, and $y_{\mathrm{in}}=\eta^{1/s}$, with $s=3$ for $7/3$ and $s=2$ for $3/2$. The dimensionless enclosed mass is
\begin{subequations}\label{eq:dimensionlessmass}
	\begin{eqnarray}
		\mu_g(y)\equiv\frac{m(r_t y^s)}{r_t}=\mu_{0,g}+\varepsilon_g H_g(y),
	\end{eqnarray}	
	with
	\begin{eqnarray}
		\varepsilon_g&=&\mathcal N_{\mathrm{sp}}\mathcal A_\rho r_t^{2-g},\\
		\mu_{0,g}&=&\frac{M_\bullet}{r_t}-\varepsilon_g H_g(y_{\mathrm{in}}),
	\end{eqnarray}
\end{subequations}
where $g$ denotes the chosen value of $\gamma_{\mathrm{sp}}$, with all model parameters evaluated for that index. Here $H_g$ is a dimensionless mass primitive, distinct from $F_g(r)$ in Eq.~\eqref{primitive}. For the benchmark,
\begin{equation}\label{eq:HsevenThirds}
	\begin{aligned}
		H_{7/3}(y)={}&-\frac{3\eta^2}{4y^4}+\frac{6(\eta+\eta^2)}y
		+\frac32(1+4\eta+\eta^2)y^2\\
		&-\frac65(1+\eta)y^5+\frac38y^8.
	\end{aligned}
\end{equation}
The shallower index gives
\begin{equation}\label{eq:HthreeHalves}
	\begin{aligned}
		H_{3/2}(y)={}&-\frac{2\eta^2}y-4(\eta+\eta^2)y
		+\frac23(1+4\eta+\eta^2)y^3\\
		&-\frac45(1+\eta)y^5+\frac27y^7.
	\end{aligned}
\end{equation}
Both follow directly by integrating the squared-taper density. Exterior mass conservation equivalently fixes $\varepsilon_g=(M_{\mathrm{ext}}/r_t)/[H_g(1)-H_g(y_{\mathrm{in}})]$.

To integrate the metric, define $P_g(y)=y^q[y^s-2\mu_g(y)]$. For $7/3$ take $q=4$, giving a degree-$12$ polynomial; for $3/2$ take $q=1$, giving a degree-$8$ polynomial. The rationalized integrand obeys
\begin{equation}\label{eq:rationalmetric}
	\frac{2m(r)}{r[r-2m(r)]}\,dr
	=\left[\frac{s y^{s+q-1}}{P_g(y)}-\frac{s}{y}\right]dy.
\end{equation}
For simple roots $z_{g,k}$ of $P_g$, partial fractions give $y^{s+q-1}/P_g(y)=\sum_{k=1}^{d_g}z_{g,k}^{s+q-1}/[P_g'(z_{g,k})(y-z_{g,k})]$, where $d_{7/3}=12$ and $d_{3/2}=8$. A prime on $P_g$ denotes differentiation with respect to its dimensionless argument. Introduce the boundary-normalized logarithmic sum
\begin{equation}\label{eq:rootmetricSum}
	\mathcal S_g(y)=\sum_{k=1}^{d_g}
	\frac{z_{g,k}^{s+q-1}}{P_g'(z_{g,k})}
	\ln\left(\frac{y-z_{g,k}}{1-z_{g,k}}\right).
\end{equation}
Using the outer boundary value $B_{III}(r_t)=1-2M_{\mathrm{tot}}/r_t$, the two explicit matter-region solutions are

\begin{subequations}\label{eq:rootmetricSolutions}
	\begin{eqnarray}
		\hspace*{-1cm}A_H^{(7/3)}(r)&=&B_{III}(r_t) y^{-3}e^{3\mathcal S_{7/3}(y)},\,\,\,
		y=(r/r_t)^{1/3},\\
		\hspace*{-1cm}A_H^{(3/2)}(r)&=&B_{III}(r_t) y^{-2}e^{2\mathcal S_{3/2}(y)},\,\,\,
		y=(r/r_t)^{1/2}.
	\end{eqnarray}
\end{subequations}
The superscripts label different density indices, not different regions. Complex-conjugate roots are paired, and the logarithms are continued from $y=1$, where each is zero. In a static occupied interval $P_g(y)\ne0$, this prescription gives the real positive solution of Eq.~\eqref{Ahalo}. Repeated roots require the corresponding repeated-pole partial fractions; Eq.~\eqref{Ahalo} remains valid in such limiting cases. The $3/2$ case is simpler because its rationalization involves eight roots rather than twelve, but neither expression generally reduces to a short formula with explicitly solvable roots.

The boundary values follow directly. At $r=r_t$, $y=1$ and $\mathcal S_g(1)=0$, so $A_H(r_t)=B_{III}(r_t)=A_{III}(r_t)$. At the inner edge,
\begin{eqnarray}\label{eq:rootmetricInner}
	A_H(r_{\mathrm{in}})&=&B_{III}(r_t) y_{\mathrm{in}}^{-s}
	e^{s\mathcal S_g(y_{\mathrm{in}})},\\
	C&=&\frac{B_{III}(r_t) y_{\mathrm{in}}^{-s}
		e^{s\mathcal S_g(y_{\mathrm{in}})}}{1-r_h/r_{\mathrm{in}}}.
\end{eqnarray}
The second line of Eq.~\eqref{eq:rootmetricInner} explicitly evaluates the same normalization constant $C$ defined by the general integral in Eq.~\eqref{eq:gapNormalization}; it introduces no additional constant. Thus $A_I(r_{\mathrm{in}})=C(1-r_h/r_{\mathrm{in}})=A_H(r_{\mathrm{in}})$. Both formulas evaluate the metric fixed by Eq.~\eqref{Ahalo}; derivative matching follows below from the field equations.

For the benchmark inputs of Section~\ref{sec:density}, the root-based expressions for both indices agree with independent high-precision quadrature at the two boundaries and two interior sample radii to relative differences below $10^{-50}$. The mass-primitive derivatives were independently checked as well. This is a numerical consistency check of the analytical representation, not an estimate of astrophysical accuracy.

The function $A(r)$ directly determines clocks and gravitational redshift: a static observer has $d\tau=\sqrt{A(r)}\,dt$, and a photon emitted at radius $r$ and received at infinity has $1+z(r)=A(r)^{-1/2}$. We state the matching conditions in terms of $A(r)$ and $B(r)$; the lapse follows from its definition above. Positive density in the static solution gives $0<C<1$.

For the benchmark, the regional metric uses $C=0.999942319131$, $M_\bullet=1.000071141M_{\mathrm{Sch}}$, and $M_{\mathrm{tot}}=1.01M_{\mathrm{Sch}}$. Region II is evaluated from Eq.~\eqref{Ahalo} with the exact mass primitive. An independent evaluation of the redshift-phase integral reproduces $C$ with an absolute residual below $10^{-11}$.

The boundary values established above also match in their radial derivatives. With $[f]_{r_b}=f(r_b^+)-f(r_b^-)$, both interfaces satisfy
\begin{equation}\label{eq:metricmatching}
	\begin{gathered}
		[A]_{r_b}=0,\quad [B]_{r_b}=0,\quad [A']_{r_b}=0,\quad [B']_{r_b}=0,\\
		r_b=r_{\mathrm{in}},r_t.
	\end{gathered}
\end{equation}
The common values are
\begin{subequations}
	\begin{eqnarray}
		(A,B)_{r_b}&=&(C_b f_b,f_b),\\
		(A',B')_{r_b}&=&\frac{2M_b}{r_b^2}(C_b,1),
	\end{eqnarray}
\end{subequations}
where $f_b=1-2M_b/r_b$, $(M_b,C_b)=(M_\bullet,C)$ at $r_{\mathrm{in}}$ and $(M_{\mathrm{tot}},1)$ at $r_t$. Since $A(r_b)>0$, the lapse, redshift, and their first derivatives also match. Equation~\eqref{eq:massedges} ensures continuity of $A''$ and $B''$. The continuity of $B'$ is an additional property of the tapered profile, beyond the shell-free junction requirements discussed below.

For the factorization $A(r)=(1-r_h/r)e^{\Gamma(r)}$, the regional restrictions of the redshift phase are
\begin{subequations}
	\begin{eqnarray}
		\Gamma_I(r)&=&\ln C,\\
		\nonumber
		\Gamma_{II}(r)&=&\ln\frac{r_t-2M_{\mathrm{tot}}}{r_t-r_h}\\
		&&\quad-\int_r^{r_t}\frac{2h(x)\,dx}{(x-r_h)[x-2m(x)]},\\
		\Gamma_{III}(r)&=&\ln\frac{r-2M_{\mathrm{tot}}}{r-r_h}.
	\end{eqnarray}
\end{subequations}
Thus $\Gamma_I(\ri)=\Gamma_{II}(\ri)=\ln C$ and $\Gamma_{II}(\rt)=\Gamma_{III}(\rt)=\ln[(\rt-2\Mt)/(\rt-\rh)]$. From both sides $\Gamma'(\ri)=0$ and $\Gamma'(\rt)=2\Me/[(\rt-\rh)(\rt-2\Mt)]$.

The absence of a surface layer is determined by continuity of the induced metric and extrinsic curvature, rather than by density continuity alone~\cite{Israel1966}. For a static timelike interface at fixed $r=r_b$, with the same normalized time coordinate on both sides, the relevant extrinsic-curvature components are proportional to $\sqrt{B(r_b)}A'(r_b)/A(r_b)$ and $\sqrt{B(r_b)}/r_b$. The matches of $A(r)$, $B(r)$, and $A'(r)$ established above therefore satisfy the shell-free junction conditions.

Density regularity supplies a stronger, separate property. Since $m'(r)=4\pi r^2\rho(r)$, a bounded jump in $\rho(r)$ can leave $m(r)$ continuous while producing a jump in $m'(r)$; the mass in an interval shrinking onto the interface tends to zero. Continuity of $\rho(r)$ makes $m(r)$ continuously differentiable, and continuity of $\rho'(r)$ additionally makes $m(r)$ twice continuously differentiable at a finite-radius interface. In the present $p_r(r)=0$ construction, the radial field equation fixes $A'(r)/A(r)$ from $m(r)$, so a finite density jump need not obstruct shell-free matching. Our squared tapers instead enforce the additional density regularity demonstrated by Eq.~\eqref{eq:edgeexpansion}. Thus smoothness is a controlled feature of this prescription, not a universal requirement that invalidates sharp-edged Einstein clusters. The outer geometry is asymptotically flat Schwarzschild. A vacuum continuation below $\rh$ has timelike $r$, so static-observer lapse and redshift interpretations apply only outside the horizon.

%%%%%%%%%%%%%%%
%%%%%%%%%%%%%%%
%%%%%%%%%%%%%%%

\section{Energy conditions and radial stability}\label{sec:admissibility}
Regions I and III are vacuum, so all standard matter energy conditions hold there automatically. The nontrivial tests below concern region II, $r_{\mathrm{in}}<r<r_t$, where $\rho_{II}(r)>0$. We specialize the energy-condition tests for black hole--halo systems~\cite{Shen2025} to the stresses in Eq.~\eqref{eq:pt}. For $\rho_{II}(r)\geq0$, $m_{II}(r)\geq0$, and $r>2m_{II}(r)$, Eq.~\eqref{eq:pt} gives $p_{t,II}(r)\geq0$. Hence the null conditions $\rho_{II}+p_{r,II}\geq0$ and $\rho_{II}+p_{t,II}\geq0$, the weak condition $\rho_{II}\geq0$, and the strong condition $\rho_{II}+p_{r,II}+2p_{t,II}\geq0$ hold. In the occupied region $r_{\mathrm{in}}<r<r_t$, where $\rho_{II}(r)>0$, the remaining dominant-energy requirement is $\rho_{II}(r)\geq p_{t,II}(r)$. Substituting Eq.~\eqref{eq:pt} gives the compactness bound
\begin{equation}
	\frac{m_{II}(r)}{r}\leq\frac{2}{5},\qquad r_{\mathrm{in}}<r<r_t.
\end{equation}
This reduction uses $\rho_{II}(r)>0$ in the occupied interval. At both edges the stresses tend to zero; the bound does not restrict the inner vacuum gap.

For a circular orbit, $\Omega_{II}(r)=d\varphi/dt$ is the coordinate angular frequency, as in the Einstein-cluster geodesic analysis~\cite{Maeda2025}. A static observer measures a proper circumferential distance $r\,d\varphi$ during a proper time $\sqrt{A_{II}(r)}\,dt$, giving the local speed $v_{II}(r)=r\Omega_{II}(r)/\sqrt{A_{II}(r)}$. The circular geodesic condition gives $\Omega_{II}^2(r)=A_{II}'(r)/(2r)$; substituting the radial field equation then yields the speed below. The resulting local speed is
\begin{equation}
	v_{II}^2(r)=\frac{rA_{II}'(r)}{2A_{II}(r)}=\frac{m_{II}(r)}{r-2m_{II}(r)}.
\end{equation}
Massive constituents require $v_{II}^2(r)<1$, which is equivalent to $r>3m_{II}(r)$. For circular constituents, the fixed-background radial stability criterion can be written as~\cite{Maeda2025}
\begin{equation}\label{stability}
	S_{II}(r)=r^2m_{II}'(r)+rm_{II}(r)-6m_{II}(r)^2>0.
\end{equation}
The gap excludes the near-horizon radii at which the assumed massive circular constituents cannot provide stable orbital support. At a smooth onset, the density and mass derivative vanish, so the limiting orbital-stability condition reduces to its Schwarzschild form at the central mass. This motivates a physical inner edge, while neither the precise depletion history nor the squared taper follows from that condition. The inner limiting conditions, consistent with Refs.~\cite{Shen2025,Maeda2025}, are $\kappa\geq5/4$ for the dominant condition, $\kappa>3/2$ for timelike circular orbits, and $\kappa\geq3$ for nonnegative radial stability at the edge. At $\kappa=3$ the limiting orbit is marginal, while occupied orbits at larger $r$ may be strictly stable. At the outer edge stability requires $\rt>6\Mt$.

A useful sufficient analytic test follows from
\begin{equation}
	\begin{aligned}
		r m_{II}'(r)&=\NN\mathcal A_\rho r^{3-\gs}(1-\ri/r)^2(1-r/\rt)^2\\
		&\leq\NN\mathcal A_\rho\rt^{3-\gs}.
	\end{aligned}
\end{equation}
If
\begin{equation}\label{sufficient}
	\NN\mathcal A_\rho\rt^{3-\gs}<\Mb,
\end{equation}
then $r m_{II}'(r)<m_{II}(r)$ and $m_{II}(r)/r$ decreases throughout the spike. For $\kappa\geq3$, this proves $m_{II}(r)/r\leq1/(2\kappa)\leq1/6$. Hence staticity, all standard energy conditions, timelike constituent orbits, and $S_{II}(r)>0$ follow throughout the open matter interval. At $\kappa=3$, only the limiting inner orbit is marginal. It also excludes a halo light ring. This criterion is sufficient, not necessary: configurations that fail it require direct evaluation of the local staticity, energy, and orbital-stability conditions. The proof concerns constituent orbits, not collective collisionless stability or the existence of a regular finite-width distribution function.

For the inputs in Section~\ref{sec:density}, the sufficient bound reads
$\mathcal N_{\mathrm{sp}}\mathcal A_\rho r_t^{3-\gamma_{\mathrm{sp}}}=0.015675088<M_\bullet=1.000071141$.
Thus $m_{II}(r)/r$ decreases from $1/6$ at $r_{\mathrm{in}}$ to $0.000101$ at $r_t$. The energy, staticity and constituent-orbit conclusions follow throughout the occupied interval, rather than from grid sampling alone.

%%%%%%%%%%%%%%%
%%%%%%%%%%%%%%%
%%%%%%%%%%%%%%%

\section{Circular orbits, shadows, and profile comparisons}\label{sec:geodesics}

We use the regional metrics defined in Section~\ref{sec:metric} for both null and timelike geodesics. In the equatorial plane, the conserved Killing energy and angular momentum give
\begin{eqnarray}\label{eq:regionalgeodesic}
	E&=&A_i(r)\dot t,\qquad L=r^2\dot\varphi,\\
	\dot r^2&=&B_i(r)\left[\frac{E^2}{A_i(r)}-\epsilon-\frac{L^2}{r^2}\right].
\end{eqnarray}
Here $\epsilon=1$ for timelike motion and $\epsilon=0$ for null motion. The dot denotes differentiation with respect to proper time for a massive particle, with $E$ and $L$ per unit rest mass, or an affine parameter for a photon. The same $E$ and $L$ remain conserved when a trajectory crosses either interface because the time coordinate is globally normalized.

The regional potential is $V_{\epsilon,i}(r;L)=A_i(r)[\epsilon+L^2/r^2]$: timelike and null potentials differ through $\epsilon$, not through a different choice of geometry. Writing $V_\epsilon(r;L)$ without $i$ below denotes this single piecewise function. The trajectory determines which pieces are sampled. An escaping scattering ray with an allowed turning point in the gap traverses all three regions; one turning in the spike traverses regions II and III, and one turning outside $r_t$ remains in region III. For a bound massive orbit, both radial turning points determine the visited regions. In the benchmark, the circular photon orbit is in region I, the marginal timelike orbit lies at the I--II interface, and the occupied dark matter circular orbits lie in region II.

For clarity, write the regional masses as $m_I(r)=M_\bullet$, $m_{II}(r)=M_\bullet+h(r)$, and $m_{III}(r)=M_{\mathrm{tot}}$. The regional domains and lapse convention are those of Section~\ref{sec:metric}. Thus the circular-geodesic relations used in Einstein-cluster analyses~\cite{Bohmer2007,Maeda2025} give, for timelike motion in region $i$,
\begin{eqnarray}
	E_{c,i}^2(r)&=A_i(r)\frac{r-2m_i(r)}{r-3m_i(r)},\\
	L_{c,i}^2(r)&=\frac{m_i(r)r^2}{r-3m_i(r)},\\
	\Omega_{c,i}^2(r)&=\frac{A_i'(r)}{2r}=\frac{A_i(r)m_i(r)}{r^2[r-2m_i(r)]}.
\end{eqnarray}
At fixed particle $L$, the potential curvature is
\begin{equation}
	V_{1,i}''(r_c;L_{c,i}(r_c))=\frac{2A_i(r_c)S_i(r_c)}{r_c^2[r_c-2m_i(r_c)][r_c-3m_i(r_c)]}.
\end{equation}
Here $S_i(r)=r^2m_i'(r)+r m_i(r)-6m_i^2(r)$, and primes denote $d/dr$. These expressions use the same region as the circular radius $r_c$; at an interface they are interpreted through their matching one-sided limits. The sign of the potential curvature distinguishes stable, unstable, and marginal circular orbits. In a Schwarzschild vacuum, marginal stability occurs at $6M$ and marginal binding at $4M$; the two conditions are distinct. In the gap, the constant redshift factor $C$ changes the energy measured at infinity, so the marginally bound radius need not equal $4M_\bullet$; this does not change the radial-stability edge at $6M_\bullet$.

The null circular-orbit condition is $r_c=3m(r_c)$ for the present $p_r(r)=0$ metric. Using the critical-impact-parameter construction for spherical shadows~\cite{Nampalliwar2021,Daghigh2024}, the gap solution gives
\begin{equation}
	r_{\mathrm{ph}}=3\Mb,\qquad b_{\mathrm{crit}}=\frac{3\sqrt3\Mb}{\sqrt C},
\end{equation}
provided it lies in the gap and controls capture. A formal matter-region light ring would satisfy $r=3[\Mb+h(r)]$; the timelike-constituent condition $r>3m(r)$ excludes such an orbit within an admissible occupied cluster. The formal exterior candidate $3\Mt$ counts only if it lies at or beyond $\rt$. For the benchmark, $r_{\mathrm{ph}}=3.000213424M_{\mathrm{Sch}}$ and $b_{\mathrm{crit}}=5.196671960M_{\mathrm{Sch}}$. The formal outer candidate $3M_{\mathrm{tot}}=3.03M_{\mathrm{Sch}}$ lies below $r_t$ and is not in region III; the only exterior light ring is the gap photon sphere.

For comparison with a Schwarzschild shadow under different mass calibrations, define the dimensionless ratios
\begin{equation}
	\mathcal R_{\mathrm{bare}}=\frac{b_{\mathrm{crit}}}{3\sqrt3\Ms},\qquad
	\mathcal R_{\mathrm{dyn}}=\frac{b_{\mathrm{crit}}}{3\sqrt3 M_{\mathrm{dyn}}}.
\end{equation}
The first uses the original black hole mass; the second uses the dynamical mass $M_{\mathrm{dyn}}$ inferred from tracer orbital motion. Here ``dynamical'' describes the mass measurement, not time dependence of the static metric. For tracers entirely outside $\rt$, $M_{\mathrm{dyn}}=\Mt$. Their different normalizations can give different apparent shadow trends. These ratios are theoretical predictions, not observational constraints by themselves. For a distant observer at distance $D$, the angular shadow radius is approximately $\theta_{\mathrm{sh}}=b_{\mathrm{crit}}/D$ and the angular diameter is $d_{\mathrm{sh}}=2b_{\mathrm{crit}}/D$. Thus a calibrated angular-size and dynamical-mass measurement would give $\mathcal R_{\mathrm{dyn}}\simeq Dd_{\mathrm{sh}}/(6\sqrt3M_{\mathrm{dyn}})$ in geometrized units. At a finite static observer radius $r_o$ outside the matter, the exact local relation is $\sin\theta_{\mathrm{sh}}=b_{\mathrm{crit}}\sqrt{A(r_o)}/r_o$. Observational use consequently requires a distance, an appropriate mass estimate, and uncertainties. The bare mass $M_{\mathrm{Sch}}$ is a bookkeeping input and is not generally measured independently. Moreover, a bright emission-ring diameter cannot simply be identified with the capture-shadow diameter without an emission model; spike-related optical modeling illustrates this distinction~\cite{Nampalliwar2021}. No observational bound or fit is imposed in the present benchmark.

The benchmark gives $\mathcal R_{\mathrm{bare}}=1.000099985$ and $\mathcal R_{\mathrm{dyn}}=0.990198005$ when $M_{\mathrm{dyn}}=M_{\mathrm{tot}}$.
The shadow enlargement is about $0.0100\%$ relative to the bare-mass Schwarzschild scale, whereas the shadow is smaller than that of a Schwarzschild hole with the same total mass. Relative to the same central mass $M_\bullet$, its enlargement is $100(C^{-1/2}-1)=0.002884\%$. These distinct calibrations explain why a steep slope does not by itself imply a large optical departure.

The influence of the density index on the same observables is shown in Table~\ref{tab:sensitivity}, at fixed reference mass, cutoff, and $\kappa$.
\begin{table*}[!t]\centering\small
	\caption{Sensitivity to the reference spike index at fixed $M_H$, $r_t$, $\kappa$, and $M_{\mathrm{Sch}}$; masses and lengths are in $M_{\mathrm{Sch}}$ units.}\label{tab:sensitivity}
	\begin{tabular}{rrrrr}\toprule $\gamma_{\mathrm{sp}}$ & $M_\bullet$ & $\NN$ & $C$ & $b_{\mathrm{crit}}$\\\midrule
		1.500000 & 1.000000147 & 4.409764 & 0.999987427 & 5.196185854\\
		2.250000 & 1.000038338 & 2.512301 & 0.999952337 & 5.196475472\\
		2.333333 & 1.000071141 & 2.351263 & 0.999942319 & 5.196671960\\
		2.500000 & 1.000244979 & 2.070563 & 0.999915540 & 5.197644872\\
		\bottomrule\end{tabular}
\end{table*}
For these inputs, increasing $\gamma_{\mathrm{sp}}$ increases the mass assigned to the central hole, decreases the gap factor $C$, and enlarges $b_{\mathrm{crit}}$. The $3/2$ row is a shallower alternative relevant to some cored initial distributions~\cite{Gondolo1999}; all rows are illustrative sensitivity examples.

\begin{figure*}[!t]\centering
	\includegraphics[width=\linewidth]{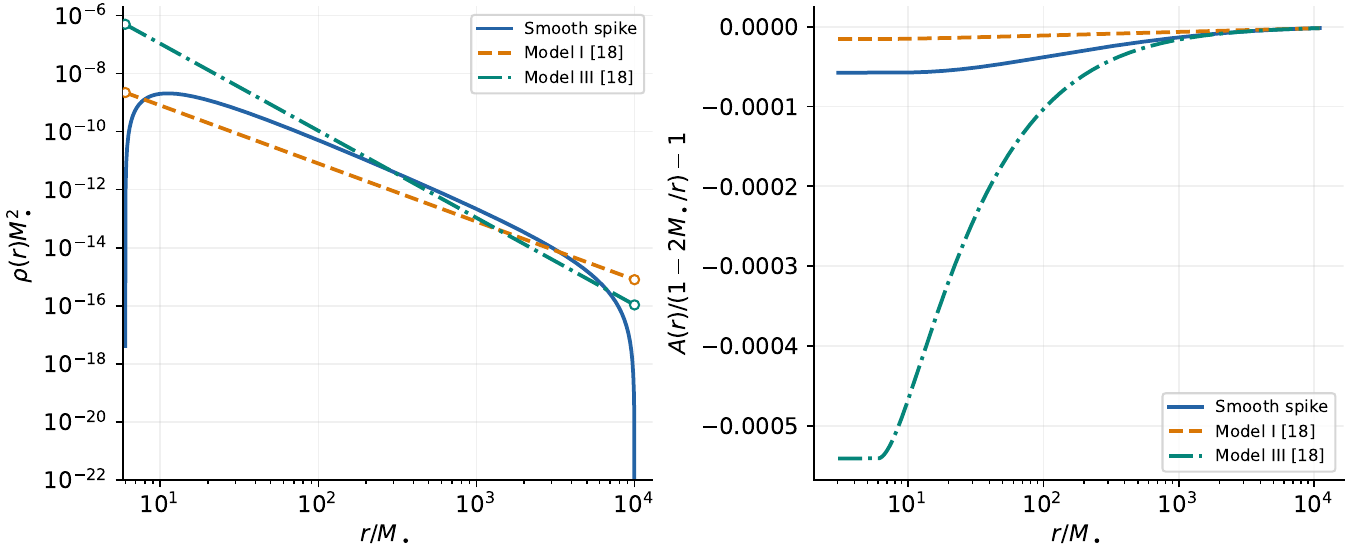}
	\caption{Density and environmental metric response for the present spike and models I and III of Ref.~\cite{r93}, at the common mass budget $M_{\mathrm{ext}}/M_\bullet=0.009928152459$ and outer radius $r_t/M_\bullet=9999.288638208$. Left: density; open circles mark finite matter-side endpoint values, while zero vacuum density is omitted on logarithmic axes. Right: fractional metric change $\delta_A(r)=A(r)/(1-2M_\bullet/r)-1$, with time normalized at infinity. Greater distance below zero indicates a stronger redshift modification. The legend labels ``Model I [18]'' and ``Model III [18]'' refer to the models of Ref.~\cite{r93}. The reference models are recalibrated rather than reproduced at their original illustrative parameters.}\label{fig:profilecomparison}
\end{figure*}

Figure~\ref{fig:profilecomparison} places the benchmark alongside models I and III of Ref.~\cite{Maeda2025}, recalibrated to the same central mass, exterior dark matter mass, outer radius, and time normalization at infinity. Their inner radii follow each model's marginal-support condition. The density panel displays the different radial distributions. The right panel plots the fractional change $\delta_A(r)=A(r)/A_{\mathrm{Sch}}(r)-1$, where $A_{\mathrm{Sch}}(r)=1-2M_\bullet/r$ is the vacuum reference at the same central mass. Zero denotes agreement with this reference; negative values correspond to a smaller $A(r)$ and a larger gravitational redshift to infinity at the same radius. In the gap, $\delta_A(r)=C-1$. Model I of Ref.~\cite{Maeda2025} produces the smallest departure for this benchmark, model III the largest, and the present spike lies between them. This ordering measures the environmental response, not observational agreement or physical superiority.

Because the profiles differ in concentration and boundary behavior, the comparison does not isolate smoothing from mass redistribution. The vanishing inner density of the present profile gives a marginal support edge at $6M_\bullet$, consistent with the smooth-onset criterion of Ref.~\cite{Maeda2025}.

The benchmark describes an isolated, finitely supported environment rather than a complete galactic halo. Its outer radius is an input; embedding the spike in a nonzero surrounding halo would require new matching and a recalculated global redshift. Relativistic capture distributions motivate inner suppression~\cite{Sadeghian2013,Ferrer2017}, while stellar heating and annihilation modify evolving spikes~\cite{Bertone2005b,Shapiro2016}; these processes do not by themselves derive the squared tapers used here. Likewise, constituent-orbit stability is distinct from collective Einstein-cluster stability~\cite{Maeda2025,Geralico2012}.

%%%%%%%%%%%%%%%
%%%%%%%%%%%%%%%
%%%%%%%%%%%%%%%

\section{Weak, strong, and regional deflection}\label{sec:lensing}
\subsection{Exact deflection and the three path intervals}
For a scattering ray with outer turning point $r_0$, the conserved impact parameter $b=L/E$ obeys $b=r_0/\sqrt{A_{i_0}(r_0)}$, where $i_0$ contains the turning point. Let $\mathcal D_i$ denote the radial domain of region $i$ specified above and set $\mathcal I_i(r_0)=\mathcal D_i\cap[r_0,\infty)$. The standard spherical-metric bending integral~\cite{Keeton2005,Bozza2002}, split over those path intervals, gives
\begin{subequations}\label{eq:exactdefl}
	\begin{eqnarray}
		\widehat\alpha(b)&=&2\sum_{i=I}^{III}\int_{\mathcal I_i(r_0)}F_i(r;b)\,dr-\pi,\\
		F_i(r;b)&=&\frac{b\sqrt{A_i(r)/B_i(r)}}{r^2\sqrt{1-A_i(r)b^2/r^2}}.
	\end{eqnarray}
\end{subequations}
Each $F_i(r;b)$ uses the metric functions of its own region. The impact parameter $b=L/E$ is conserved along the complete ray. The regional azimuthal changes can be written uniformly as
\begin{equation}\label{eq:regionalazimuth}
	\Delta\varphi_i(b)=2\int_{\mathcal I_i(r_0)}F_i(r;b)\,dr.
\end{equation}
An untraversed region contributes zero, and the first traversed interval begins at $r_0$. This expression covers turning points in any of the three regions. The factor two includes incoming and outgoing motion. Their sum is the total azimuthal change, so $\widehat\alpha(b)=\sum_i\Delta\varphi_i(b)-\pi$.

The swept angles $\Delta\varphi_i(b)$ include the ordinary coordinate change of a straight ray. To define additive bending contributions, we subtract the flat-space angles for a reference ray with the same closest radius $r_0$ and the same radial boundaries:
\begin{equation}\label{eq:flatreference}
	f_i(r_0)=2\int_{\mathcal I_i(r_0)}\frac{r_0\,dr}{r\sqrt{r^2-r_0^2}},\qquad
	\sum_{i=I}^{III}f_i(r_0)=\pi.
\end{equation}
The intervals $\mathcal I_i(r_0)$ are the same traversed radial intervals as in Eq.~\eqref{eq:exactdefl}; an empty interval contributes zero. The factor two includes the incoming and outgoing branches. The $f_i(r_0)$ are reference coordinate angles, not flat-space deflections: their total is $\pi$, so the total bending in flat space is zero. The regional bending contributions are then
\begin{equation}\label{eq:regionalshare}
	\alpha_i^{\mathrm{part}}(b;r_0)=\Delta\varphi_i(b)-f_i(r_0),\qquad
	\sum_i\alpha_i^{\mathrm{part}}(b;r_0)=\widehat\alpha(b).
\end{equation}
The physical ray has $b=r_0/\sqrt{A_{i_0}(r_0)}$, whereas the reference ray has $b_{\mathrm{flat}}=r_0$; the matching condition is the closest radius, not the impact parameter. This is a path-based partition, not a unique decomposition by gravitating source. In particular, the central black hole's field is present in region II. A large region-II share is not the fraction caused by dark matter. Other reference conventions can give different shares or even negative ones; their sum remains the same asymptotic bending angle.

The exact expression in Eq.~\eqref{eq:exactdefl} applies to scattering rays in all three domains. The following weak- and strong-deflection formulas describe its respective limiting regimes.

\subsection{Weak deflection}
The impact parameter $b=L/E$ is the same conserved quantity in every traversed region; no separate regional impact parameters are introduced.
The weak regime requires $2m(r)/r\ll1$ along the ray and a small total angle. For rays entirely outside the spike ($r_0\geq\rt$), the exact result is Schwarzschild with mass $\Mt$. Its expansion in the invariant impact parameter is~\cite{Keeton2005}
\begin{equation}\label{eq:weak}
	\begin{aligned}
		\widehat\alpha_{\mathrm{weak}}(b)={}&\frac{4\Mt}{b}+\frac{15\pi}{4}\left(\frac{\Mt}{b}\right)^2\\
		&+\frac{128}{3}\left(\frac{\Mt}{b}\right)^3+O[(\Mt/b)^4].
	\end{aligned}
\end{equation}
For these rays, $\alpha_I^{\mathrm{part}}=\alpha_{II}^{\mathrm{part}}=0$ and $\alpha_{III}^{\mathrm{part}}=\widehat\alpha(b)$. A weak ray can also cross the dilute spike, but then its actual $m(r)$ and $A(r)$ must be retained. The exterior expression using $\Mt$ is not a general approximation for rays that turn within the spike.

At $r_0=100M_{\mathrm{Sch}}$, the path shares are $99.50$ percent in region II and $0.50$ percent in region III. At $r_0=1000M_{\mathrm{Sch}}$ they are $94.97$ and $5.03$ percent. At $r_0=20000M_{\mathrm{Sch}}$ only region III is traversed. These percentages do not measure the fraction produced by the spike's mass.

\subsection{Strong deflection near the gap photon sphere}
The standard strong-deflection expansion is logarithmic near an unstable photon sphere~\cite{Bozza2002}. Here it applies as $r_0\to r_{\mathrm{ph}}^+$ and $b\to b_{\mathrm{crit}}^+$. For the gap light ring,
\begin{equation}\label{eq:strong}
	\widehat\alpha_{\mathrm{strong}}(b)
	=-\overline a\ln\left(\frac{b}{b_{\mathrm{crit}}}-1\right)+\overline b+o(1),
	\qquad \overline a=1.
\end{equation}
The coefficient is one because the gap integrand maps to Schwarzschild with mass $M_\bullet$ under $b\sqrt C$. The spike affects $b_{\mathrm{crit}}$ through $C$ and changes the finite constant through the rest of the path. The Schwarzschild finite term is $\ln[216(7-4\sqrt3)]-\pi$~\cite{Bozza2002}. Adding the regular difference between our outer path and the Schwarzschild path gives
\begin{equation}
	\begin{aligned}
		\overline b={}&\ln[216(7-4\sqrt3)]-\pi\\
		&+2\sum_{i=II}^{III}\int_{\mathcal D_i}\left[F_i(r;b_{\mathrm{crit}})-F_{\mathrm{Sch}}(r;b_S)\right]dr.
	\end{aligned}
\end{equation}
where $b_S=3\sqrt3M_\bullet$ and
\begin{equation}
	F_{\mathrm{Sch}}(r;b_S)=
	\frac{b_S}{r^2\sqrt{1-(1-2M_\bullet/r)b_S^2/r^2}}.
\end{equation}
The difference integral is regular because the lower endpoint lies outside the light ring. For the benchmark, $\overline b=-0.400222954$. Direct near-critical evaluations converge to this value. The logarithmic divergence is accumulated in region I, while the region-II and region-III path terms tend to finite limits.

At $r_0=1.02r_{\mathrm{ph}}$ the gap accounts for $87.53$ percent, region II for $12.47$ percent, and region III for $0.0061$ percent under Eq.~\eqref{eq:regionalshare}. At $1.001r_{\mathrm{ph}}$ the gap share reaches $93.06$ percent and tends to one hundred percent in the logarithmic limit. This does not remove the global spike dependence of $b_{\mathrm{crit}}$ and $\overline b$.

The total deflection remains continuous as the turning point crosses $r_{\mathrm{in}}$, while the gap contribution tends to zero as $r_0\to r_{\mathrm{in}}^-$. For the benchmark ray with $r_0=r_{\mathrm{in}}$, the angle is $1.014886$ rad, about $1.04\times10^{-5}$ rad larger than for Schwarzschild at the same central mass and turning radius. This difference measures the effect of adding the spike; it is distinct from the region-II path share. The ray is intermediate rather than near-critical because $r_{\mathrm{in}}=2r_{\mathrm{ph}}$.

\begin{figure*}[!t]
	\centering
	\includegraphics[width=\linewidth]{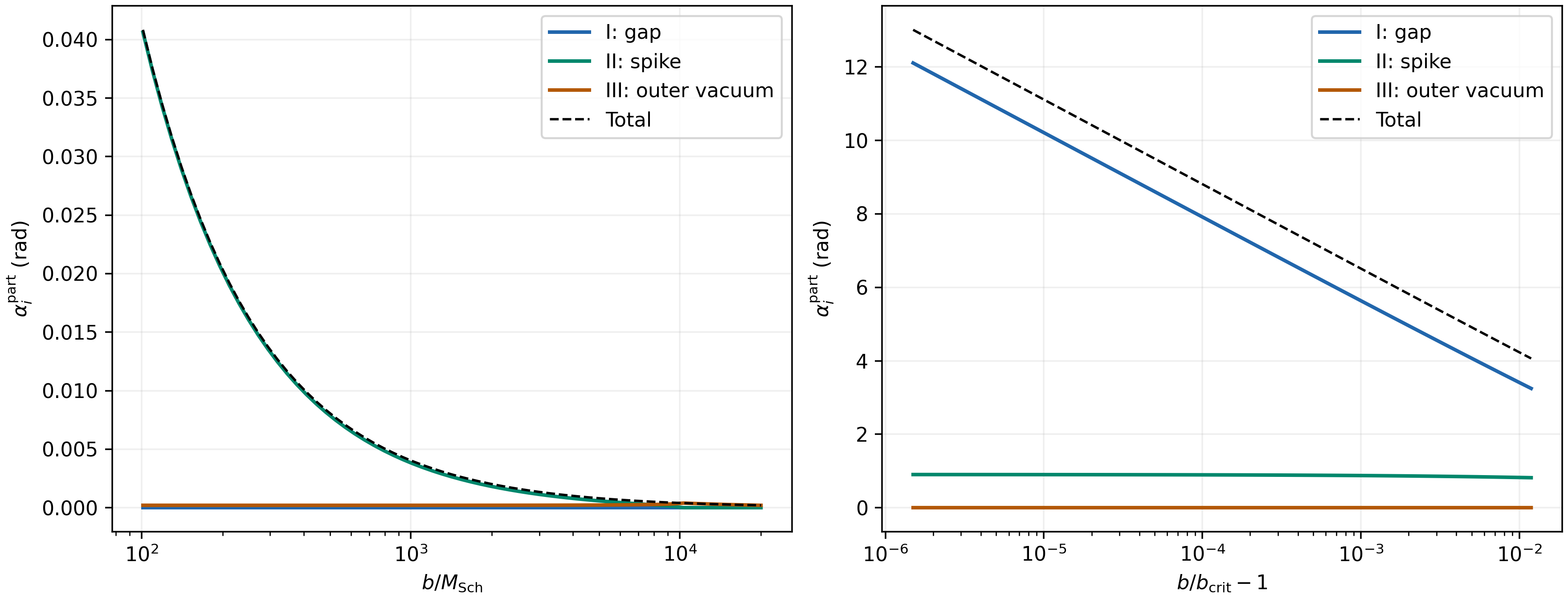}
	\caption{Path-based bending contributions defined by Eq.~\eqref{eq:regionalshare}: weak rays (left) and strong near-critical rays (right). The dashed curve is their sum. These are radial-path shares, not black hole versus spike source fractions.}\label{fig:regional}
\end{figure*}

Figure~\ref{fig:regional} compares the regional contributions for weak rays and rays approaching the gap photon sphere.

%%%%%%%%%%%%%%%
%%%%%%%%%%%%%%%
%%%%%%%%%%%%%%%

\section{Photon-orbit scales and scalar-wave propagation}\label{sec:waves}

The orbital calculation supplies two complementary time scales, with the same Killing time normalized at infinity. The coordinate angular frequency measures the photon's orbital rate, while the Lyapunov exponent measures the exponential growth rate of a small radial displacement from the unstable orbit. They characterize the leading eikonal correspondence~\cite{Cardoso2009}. At an unstable null circular orbit $r_c$ in region $i$, the metric functions are specifically $A_i(r)$ and $B_i(r)$:
\begin{eqnarray}
	\Omega_{\mathrm{ph},i}(r_c)&=&\frac{\sqrt{A_i(r_c)}}{r_c},\\
	\lambda_i^2(r_c)&=&\Omega_{\mathrm{ph},i}^2(r_c)[1-3m_i'(r_c)].
\end{eqnarray}
In region I (the gap), $\Omega_{\mathrm{ph},I}(r_c)=\lambda_I(r_c)=\sqrt C/(3\sqrt3\Mb)$. A region-II candidate uses $m_{II}'(r_c)$ and $A_{II}(r_c)$; an exterior Schwarzschild candidate would give $1/(3\sqrt3\Mt)$ only if that orbit lay in the exterior domain. The benchmark has no additional photon sphere in the matter or outer-vacuum region. Its gap frequency and Lyapunov exponent are $\Omega_{\mathrm{ph}}=\lambda=0.192430850M_{\mathrm{Sch}}^{-1}$.

Applying the photon-orbit correspondence to the test-field branch gives~\cite{Cardoso2009}
\begin{equation}
	\begin{aligned}
		\omega_{\ell n}\simeq{}&(\ell+\tfrac12)\Omega_{\mathrm{ph},i}(r_c)\\
		&-i(n+\tfrac12)\lambda_i(r_c),\quad \ell\gg1,\ n\geq0.
	\end{aligned}
\end{equation}
Here $i$ selects the region containing the relevant unstable orbit; it does not define independent spectra in the three regions. The real part describes oscillation and the negative imaginary part describes damping for the convention $e^{-i\omega t}$. In the gap, both scales contain $\sqrt C/M_\bullet$. Relative to Schwarzschild with the same central mass, both the leading oscillation and damping rates are multiplied by $\sqrt C$; this local photon-orbit statement does not imply an identical rescaling of the entire finite-$\ell$ spectrum. The benchmark has $\sqrt C\simeq0.999971159$, corresponding to a reduction of about $0.002884\%$ in these leading rates. The high-multipole approximation is a test-field diagnostic, not a calculation of the coupled gravitational and matter spectrum.

To probe the complete geometry, consider a minimally coupled massless scalar field, $\Box\Psi=0$, as in studies of matter-surrounded black holes~\cite{Leung1999,Daghigh2022}. Its separated form is $\Psi(t,r,\theta,\varphi)=e^{-i\omega t}Y_{\ell m}(\theta,\varphi)\psi_\ell(r)/r$. The spherical harmonic contributes the angular eigenvalue $\ell(\ell+1)$. Use one global tortoise coordinate $x(r)=r_*(r)$ with regional restrictions $x_i(r)$ and additive constant fixed by $x(r_{\mathrm{in}})=0$. Its explicit expressions are given in Eq.~\eqref{eq:regionaltortoise}. The separated regional wave equation is
\begin{equation}
	\frac{d^2\psi_{\ell,i}}{dx_i^2}+[\omega^2-V_{\ell,i}(r)]\psi_{\ell,i}=0.
\end{equation}
The same $\omega$ is used in all three regions, since $t$ is globally normalized. An omitted region index denotes the complete matched function. Substitution of the background field equations with $p_{r,i}=0$ gives the scalar effective potential directly in terms of the enclosed mass:
\begin{equation}\label{eq:scalarVm}
	V_{\ell,i}(r)=A_i(r)\left[\frac{\ell(\ell+1)}{r^2}+\frac{2m_i(r)}{r^3}-\frac{m_i'(r)}{r^2}\right].
\end{equation}
The first term in the bracket is the angular-momentum barrier; the remaining terms describe the enclosed-mass and local-density contributions. Unlike the geodesic potential $V_\epsilon(r;L)$, this wave potential includes the scalar angular index and the mass derivative. It applies throughout the three-region exterior using $m_I=M_\bullet$, $m_{II}(r)=M_\bullet+h(r)$, and $m_{III}=M_{\mathrm{tot}}$, with the corresponding $A_i(r)$. The matter contribution is confined to region II. Smooth metric matching makes the complete potential continuous at both interfaces, without a delta-function term.

For configurations obeying the sufficient bound in Eq.~\eqref{sufficient}, scalar-potential positivity follows analytically rather than only from a radial scan. In region II that bound gives $r m'(r)<m(r)$. The region-II specialization of Eq.~\eqref{eq:scalarVm} therefore has a positive bracket for every $\ell=0,1,\ldots$, and $A_{II}(r)>0$ implies $V_{\ell,II}(r)>0$.
The vacuum specializations of Eq.~\eqref{eq:scalarVm} are also positive for $r>r_h$, with zero limiting potential at the horizon and at infinity. This excludes exponentially growing modes of the minimally coupled massless scalar test field. Indeed, a hypothetical mode with $\operatorname{Im}\omega>0$ satisfying the ingoing horizon and outgoing infinity conditions decays at both ends of the global coordinate $x$. Multiplying its radial equation by $\psi_\ell^*$ and integrating therefore gives
\begin{equation}\label{eq:scalarEnergyIdentity}
	\int_{-\infty}^{\infty}
	\left(\left|\frac{d\psi_\ell}{dx}\right|^2+V_\ell(x)|\psi_\ell(x)|^2\right)dx
	=\omega^2\int_{-\infty}^{\infty}|\psi_\ell(x)|^2dx.
\end{equation}
The regional interface terms cancel because the wave and its tortoise derivative match. For a nonzero solution the left-hand side is real and nonnegative, requiring $\omega^2$ to be real and nonnegative. With $\operatorname{Im}\omega>0$, reality would require $\operatorname{Re}\omega=0$, which instead gives $\omega^2<0$, a contradiction. This argument concerns a hypothetical growing mode; the square-integrable identity is not being applied to the usual damped QNM wave functions. It establishes scalar test-field mode stability under the sufficient bound, not collective stability of the Einstein cluster or stability of coupled matter--metric perturbations.

\begin{figure*}[!t]\centering
	\includegraphics[width=\linewidth]{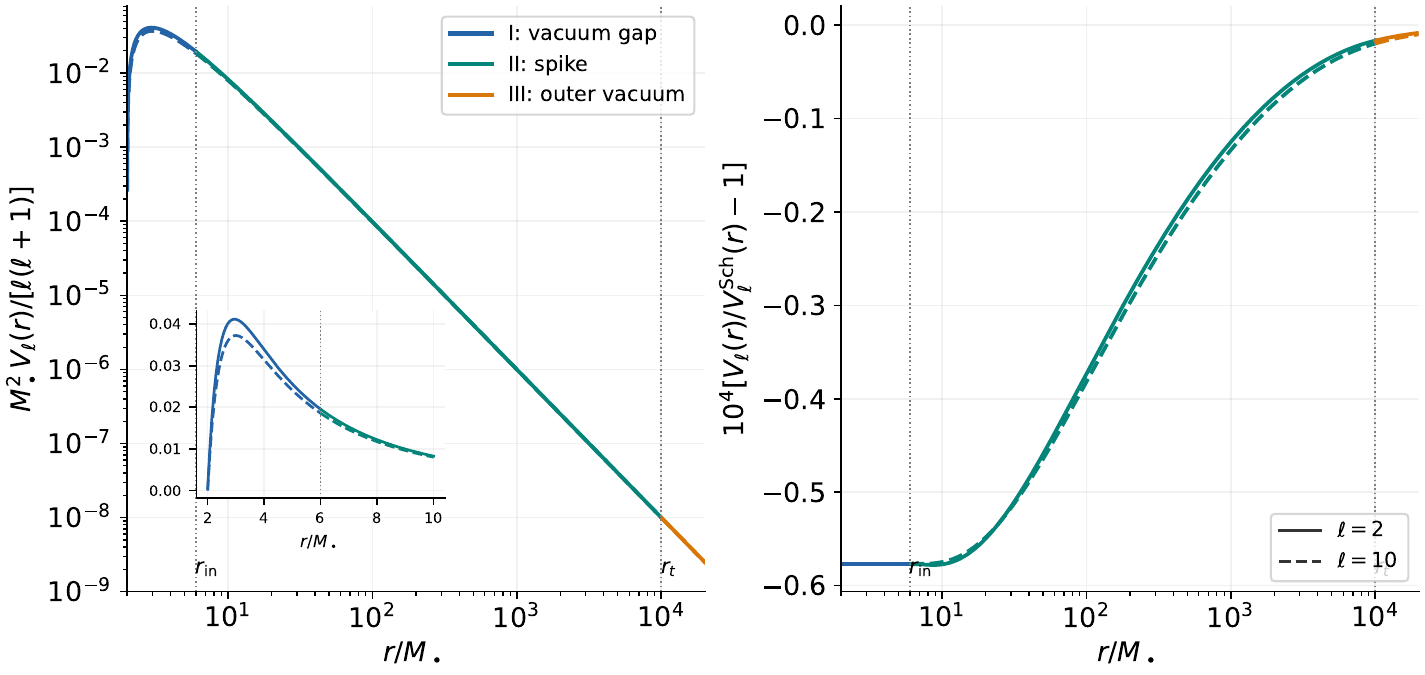}
	\caption{Massless-scalar potentials at $\ell=2$ (solid) and $10$ (dashed), colored blue in region I (vacuum gap), green in region II (spike), and orange in region III (outer vacuum). Left: normalized barrier $M_\bullet^2V_\ell(r)/[\ell(\ell+1)]$ on logarithmic axes across the full radial range; the inset resolves the near-hole barrier on linear axes. Right: $10^4[V_\ell(r)/V_\ell^{\mathrm{Sch}}(r)-1]$, with a central-mass Schwarzschild reference normalized at infinity. Dotted vertical lines mark $r_{\mathrm{in}}$ and $r_t$. The colored segments join continuously. These are potential diagnostics, not computed quasinormal frequencies or waveforms.}\label{fig:scalarpotential}
\end{figure*}

Figure~\ref{fig:scalarpotential} illustrates the barrier and its environmental modification across all three regions. Colors identify the radial regions and line styles distinguish $\ell=2$ and $10$. Dividing the potential by $\ell(\ell+1)$ makes the approach to the null-geodesic barrier visible at increasing $\ell$. The comparison with central-mass Schwarzschild uses the same infinity-normalized time and therefore retains the gap redshift. Near the hole, the fractional gap difference is exactly $C-1$; at larger radii, the changing enclosed mass and density prevent a common constant rescaling.

For the matched metric, the regional restrictions of the global tortoise coordinate have the compact form
\begin{equation}\label{eq:regionaltortoise}
	x_i(r)=x(a_i)+\int_{a_i}^{r}\frac{du}{\sqrt{A_i(u)B_i(u)}}.
\end{equation}
Choose $a_I=a_{II}=r_{\mathrm{in}}$ and $a_{III}=r_t$, with $x(r_{\mathrm{in}})=0$ and $x(r_t)=\int_{r_{\mathrm{in}}}^{r_t}[A_{II}(u)B_{II}(u)]^{-1/2}\,du$. These anchors fix the additive constants rather than introducing independent regional coordinates. The one-sided values and first derivatives match at both interfaces. Thus the single coordinate $x$ runs continuously from $-\infty$ at the horizon to $+\infty$ at spatial infinity.

The regional solutions $\psi_{\ell,i}$ form one matched wave at the same frequency $\omega$. Their one-sided limits satisfy
\begin{equation}\label{eq:wavematching}
	\begin{gathered}
		[\psi_\ell]_{r_b}=0,\qquad
		\left[\sqrt{A B}\,\frac{d\psi_\ell}{dr}\right]_{r_b}=0,\\
		r_b=r_{\mathrm{in}},r_t.
	\end{gathered}
\end{equation}
Here the brackets denote the outer-minus-inner one-sided limits of the matched regional functions. At $r_{\mathrm{in}}$ they compare regions II and I; at $r_t$ they compare regions III and II. Since $d/dx=\sqrt{A_i(r)B_i(r)}\,d/dr$ in region $i$, Eq.~\eqref{eq:wavematching} enforces continuity of the wave and its tortoise derivative. The common metric factor is $\sqrt C(1-2M_\bullet/r_{\mathrm{in}})$ at the inner interface and $1-2M_{\mathrm{tot}}/r_t$ at the outer interface. Both are positive, so the one-sided radial derivatives also agree.

These conditions follow from the locally bounded potential: integrating the wave equation across either interface preserves its tortoise derivative, while continuity of the wave excludes a distributional surface interaction.

For time dependence $e^{-i\omega t}$, the usual black hole resonance conditions~\cite{Leung1999,IyerWill1987} are imposed only at the ends of the full exterior:
\begin{subequations}\label{eq:globalqnmends}
	\begin{eqnarray}
		\psi_{\ell,I}&\sim& e^{-i\omega x_I}\quad(r\to r_h^+),\\
		\psi_{\ell,III}&\sim& e^{+i\omega x_{III}}\quad(r\to\infty).
	\end{eqnarray}
\end{subequations}
One may integrate the ingoing horizon solution and the outgoing infinity solution through their respective regions, continuing them with Eq.~\eqref{eq:wavematching}. Frequencies for which the continued solutions are linearly dependent satisfy both global end conditions. No independent ingoing or outgoing condition is imposed at $r_{\mathrm{in}}$ or $r_t$: both locally independent solutions generally contribute in the matter region. This construction defines one global scalar spectrum, not a separate spectrum for each region. We next evaluate local WKB estimates and real-frequency scattering coefficients. The exact global complex-frequency spectrum and coupled matter--metric perturbations remain separate calculations.

The global end conditions in Eq.~\eqref{eq:globalqnmends} motivate a local approximation near the potential maximum. We implement the third-order WKB prescription of Iyer and Will~\cite{IyerWill1987}, extending the leading WKB construction of Schutz and Will~\cite{Schutz1985}, using the Schwarzschild scalar problem~\cite{Iyer1987} as a reference. Higher-order extensions and practical limitations of the WKB approximation are discussed in Refs.~\cite{Konoplya2003,Konoplya2019}. We retain third order here. Table~\ref{tab:wkb0} reports representative independently calculated estimates: low- and higher-multipole fundamental modes and one overtone. Additional modes follow the same local scaling. The overtone index is $n=0,1,2,\ldots$, with $n=0$ the fundamental mode. Define $\nu=n+1/2$. We write
\begin{equation}
	V_0^{(k)}=\left.\frac{d^k V_\ell}{dx^k}\right|_{r=r_p},\qquad V_0=V_\ell(r_p),\qquad V_0''<0.
\end{equation}
Here $r_p$ is the barrier maximum, and $k$ is the derivative order. Third-order WKB requires $k\leq6$, with $V_0^{(1)}=0$. For a maximum in region $i_p$, each derivative uses $d/dx=\sqrt{A_{i_p}(r)B_{i_p}(r)}\,d/dr$. The third-order prescription is
\begin{equation}\label{eq:wkb3}
	\omega_{\ell n}^2=V_0+\sqrt{-2V_0''}\,\Lambda_{\ell n}
	-i(n+\tfrac12)\sqrt{-2V_0''}\,(1+\Omega_{\ell n}),
\end{equation}
where $\Lambda_{\ell n}$ and $\Omega_{\ell n}$ are the standard third-order correction terms given in Ref.~\cite{IyerWill1987}.
Both corrections depend on $n$ through $\nu^2=(n+1/2)^2$, as well as on $\ell$ through the potential derivatives. We select the square root with positive real part and negative imaginary part. These are scalar test-field estimates; they are not coupled gravitational--matter modes.

For the inputs of Section~\ref{sec:density}, the potential maximum lies in region I. Writing $\mathcal L=\ell(\ell+1)$, its location for $\ell\geq1$ is
\begin{equation}
	\frac{r_p}{M_\bullet}=\frac{3(\mathcal L-1)+\sqrt{9(\mathcal L-1)^2+32\mathcal L}}{2\mathcal L}<3.
\end{equation}
Here $r_{\mathrm{in}}=6M_\bullet$. A numerical radial scan of the benchmark potential at $\ell=1,\ldots,5$ finds positive, decreasing potentials outside the gap and no additional peak. Within the gap,
\begin{equation}
	\begin{aligned}
		V_\ell&=C V_\ell^{\mathrm{Sch}},\qquad
		\frac{d}{dx}=\sqrt C\,\frac{d}{dx_{\mathrm{Sch}}},\\
		V_0^{(k)}&=C^{1+k/2}V_{0,\mathrm{Sch}}^{(k)}.
	\end{aligned}
\end{equation}
The potential contributes one factor $C$ and each tortoise derivative another $\sqrt C$. The dimensionless correction terms $\Lambda_{\ell n}$ and $\Omega_{\ell n}$ are consequently unchanged, giving
\begin{equation}\label{eq:wkbscale}
	\omega_{\ell n}^{\mathrm{sp,WKB}}=\sqrt C\,\omega_{\ell n}^{\mathrm{Sch,WKB}}.
\end{equation}
The reference uses the same central mass and time normalization. It is distinct from taking $M_H=0$ at fixed $M_{\mathrm{Sch}}$, which also sets $M_{\mathrm{abs}}=0$ and changes the central mass to $M_{\mathrm{Sch}}$. For $C=0.999942319131$, both the real frequency and damping-rate magnitude decrease by $0.00288409\%$. This algebraic scaling is exact within this local WKB construction, not an exact statement about the global spectrum. Regions II and III enter its gap normalization through $C$, but their propagation effects are not independently resolved by peak derivatives. They do enter the full boundary-value problem.

\begin{table*}[t]\centering\small
	\caption{Representative scalar frequencies independently calculated at third WKB order using Refs.~\cite{IyerWill1987,Iyer1987}, with and without the benchmark spike at the same central mass and infinity-normalized time. The spike column follows Eq.~\eqref{eq:wkbscale}. The displayed digits resolve this local redshift scaling and do not indicate accuracy relative to the exact global spectrum.}\label{tab:wkb0}
	\begin{tabular}{rrrr}\toprule
		$\ell$ & $n$ & \multicolumn{1}{c}{Schwarzschild} & \multicolumn{1}{c}{With spike}\\
		& & $M_\bullet\omega_{\ell n}^{\mathrm{Sch,WKB}}$ & $M_\bullet\omega_{\ell n}^{\mathrm{sp,WKB}}$\\\midrule
		1 & 0 & $0.29111412-0.09800136i$ & $0.29110572-0.09799854i$ \\
		2 & 0 & $0.48321103-0.09680485i$ & $0.48319709-0.09680206i$ \\
		2 & 1 & $0.46319172-0.29581013i$ & $0.46317836-0.29580160i$ \\
		5 & 0 & $1.05957032-0.09633890i$ & $1.05953977-0.09633612i$ \\
		\bottomrule\end{tabular}
\end{table*}

The precision in Table~\ref{tab:wkb0} resolves the common local redshift scaling, not the accuracy of the global resonances. The WKB expansion is asymptotic, and increasing its order does not guarantee improved accuracy~\cite{Konoplya2019}. For the present third-order estimates, order sensitivity exceeds the small benchmark shift, particularly for low multipoles and higher overtones. An independent global complex-frequency calculation is needed to establish the exact spectral changes.

To examine propagation through the complete three-region geometry, we next calculate real-frequency transmission and reflection. For this scattering problem, $\omega>0$, retain the ingoing horizon condition of Eq.~\eqref{eq:globalqnmends} and the interface matching. Replace the outgoing-only infinity condition by an incident and reflected wave~\cite{Harmark2010}:
\begin{equation}\label{eq:scatterends}
	\psi_{\ell,III}\sim e^{-i\omega x_{III}}+
	\mathcal R_\ell e^{i\omega x_{III}}\quad(x_{III}\to+\infty).
\end{equation}
With incident amplitude normalized to unity, let $\mathcal T_\ell$ be the ingoing horizon amplitude. Conservation of flux gives $T_\ell=|\mathcal T_\ell|^2$, $R_\ell=|\mathcal R_\ell|^2$, and $T_\ell+R_\ell=1$. The global tortoise coordinate already includes the gap redshift, so no extra factor $C$ multiplies $T_\ell$.

The matching conditions couple the waves across both matter edges. Regional amplitudes therefore combine coherently, including their relative phases; transmission and reflection probabilities cannot be assigned independent additive shares to the three regions.

We integrate the scalar radial equation successively through regions I, II, and III, imposing the ingoing horizon condition and continuity of the wave and its tortoise derivative at both interfaces. Matching to incident and reflected asymptotic waves in the outer vacuum determines $T_\ell$ and $R_\ell$. The numerical implementation uses adaptive integration and asymptotic expansions at the finite extraction boundaries. We compare with Schwarzschild at the same central mass to isolate the environmental response. At representative points $(\ell,\omega M_\bullet)=(0,0.1),(1,0.3),(2,0.5)$, tightening integration tolerances, doubling the outer extraction radius, and varying the near-horizon cutoff or tortoise origin change $T_\ell$ by less than $2\times10^{-12}$. An independent direct wave integration at $(1,0.3)$ agrees within $6\times10^{-12}$. These checks resolve the numerical scattering response at the tested points; they do not bound the WKB approximation error or uncertainty in the assumed density profile.

At $\ell=1$ and $\omega M_\bullet=0.3$, transmission increases by $6.96898\times10^{-5}$ and reflection decreases by the same amount. The relative transmission change is $0.013334\%$. The sign and size refer to this benchmark and sampled frequencies, not a universal enhancement for every spike. A same-total-mass Schwarzschild reference instead gives $T=0.54654260$ at this frequency; that comparison changes the central horizon and must be distinguished from the central-mass reference.

\begin{figure}[tb]\centering\includegraphics[width=\linewidth]{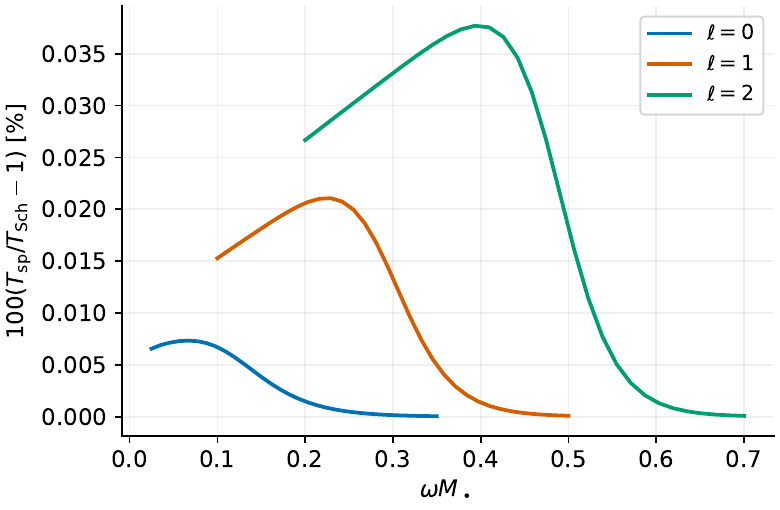}
	\caption{Relative change in scalar transmission with respect to Schwarzschild at the same central mass. Colors distinguish $\ell=0,1,2$, with 32 frequency samples per multipole. The coefficients include propagation and matching through all three regions. The enhancement shown here is specific to the benchmark and sampled frequencies.}\label{fig:scatterfreq}
\end{figure}

Figure~\ref{fig:scatterfreq} isolates the relative transmission change; the absolute probabilities nearly overlap for this benchmark. Reflection supplies no independent probability curve because flux conservation gives $R_\ell=1-T_\ell$. Colors identify the multipole $\ell$; the global coefficients are obtained by matching waves across both interfaces of the potential in Fig.~\ref{fig:scalarpotential}.

%%%%%%%%%%%%%%%
%%%%%%%%%%%%%%%
%%%%%%%%%%%%%%%

\section{Conclusions}\label{sec:conclusion}
A finite dark matter spike can retain an extended approximate power law while approaching vacuum smoothly at both prescribed boundaries. In the present density, $\rho(r)\propto r^{-\gamma_{\mathrm{sp}}}(1-r_{\mathrm{in}}/r)^2(1-r/r_t)^2$, the inner factor creates a depleted edge and a finite density maximum, whereas the outer factor truncates the distribution without changing the assigned exterior mass after normalization. The effective index $\gamma_{\mathrm{eff}}(r)=-d\ln\rho(r)/d\ln r$ quantifies the local slope within the occupied interval, distinguishing its approximate power-law portion from the rising inner density and declining outer taper. The result is a controlled equilibrium profile, rather than a prediction of the taper from a formation calculation.

The mass and metric equations provide a consistent relativistic realization, with junctions free of surface layers. For indices $7/3$ and $3/2$, rationalizing the radius yields exact metric representations in terms of polynomial roots, with their normalization and inner redshift fixed by the vacuum boundaries. The squared tapers ensure continuity of the density and its first derivative. Across the inner vacuum gap, spike, and outer vacuum, the metric functions and their first radial derivatives join continuously, as do the geodesic and massless-scalar effective potentials. A single analytic sufficient bound, together with $\kappa\geq3$, guarantees staticity, the standard energy conditions, and radial stability of the occupied constituent circular orbits; the benchmark realizes these conditions with a marginal limiting inner orbit at $6M_\bullet$.

The vacuum gap is essential to this construction: it separates the supported matter from the horizon and contains the benchmark photon sphere, while the constant factor $C$ preserves the influence of the exterior spike on clocks and strong-field observables within the gap. At a common central mass, exterior mass, and outer radius, the comparison with established Einstein-cluster profiles illustrates how different radial distributions produce different metric responses. It does not isolate boundary regularity from concentration or establish a preferred profile. The shadow comparison also depends on whether the reference mass is the central or total mass. Strong bending retains a global dependence on the spike even though its divergent path contribution arises in the vacuum gap.

The same sufficient bound guarantees a positive massless-scalar potential and excludes exponentially growing modes of this test field. Transmission probes propagation through the three matched regions, whereas the local WKB estimates inherit the gap redshift. These wave diagnostics do not establish the exact global resonance shifts or collective matter--metric stability.

The finite-mass normalization and explicit vacuum matching make this profile useful for controlled comparisons of depletion, truncation, and mass redistribution. Its astrophysical interpretation requires additional input: a formation model must determine the taper and support scales, and a coupled matter--metric analysis must establish collective stability. Boundary regularity by itself cannot select the preferred dark matter distribution.

\section*{Data availability}
No observational or experimental datasets were used in this theoretical study. The numerical results are generated from the model equations and parameter choices specified in the article; no proprietary or restricted input data are required. Published density prescriptions and analytical methods used for comparison are cited in the text.

%\acknowledgments
%acknowledgments

\bibliographystyle{mybibstyle}
\setlength{\emergencystretch}{3em}

\bibliography{bib}

@article{Bertone2005,
	author = "Bertone, Gianfranco and Hooper, Dan and Silk, Joseph",
	title = "{Particle dark matter: Evidence, candidates and constraints}",
	doi = "https://doi.org/10.1016/j.physrep.2004.08.031",
	journal = "Phys. Rep.",
	volume = "405",
	pages = "279--390",
	year = "2005"
}

@article{Nishikawa2019,
    author = "Nishikawa, H. and Kovetz, Ely D. and Kamionkowski, Marc and Silk, Joseph",
    title = "{Primordial-black-hole mergers in dark-matter spikes}",
    doi = "https://doi.org/10.1103/PhysRevD.99.043533",
    journal = "Phys. Rev. D",
    volume = "99",
    number = "4",
    pages = "043533",
    year = "2019"
}

@article{Salucci2019,
	author = "Salucci, Paolo",
	title = "{The distribution of dark matter in galaxies}",
	doi = "https://doi.org/10.1007/s00159-018-0113-1",
	journal = "Astron. Astrophys. Rev.",
	volume = "27",
	pages = "2",
	year = "2019"
}

@article{Gondolo1999,
	author = "Gondolo, Paolo and Silk, Joseph",
	title = "{Dark Matter Annihilation at the Galactic Center}",
	doi = "https://doi.org/10.1103/PhysRevLett.83.1719",
	journal = "Phys. Rev. Lett.",
	volume = "83",
	pages = "1719--1722",
	year = "1999"
}

@article{Ullio2001,
	author = "Ullio, Piero and Zhao, Hongsheng and Kamionkowski, Marc",
	title = "{Dark-matter spike at the galactic center?}",
	doi = "https://doi.org/10.1103/PhysRevD.64.043504",
	journal = "Phys. Rev. D",
	volume = "64",
	pages = "043504",
	year = "2001"
}

@article{Merritt2002,
	author = "Merritt, David and Milosavljevi{\'c}, Milo{\v{s}} and Verde, Licia and Jimenez, Raul",
	title = "{Dark Matter Spikes and Annihilation Radiation from the Galactic Center}",
	doi = "https://doi.org/10.1103/PhysRevLett.88.191301",
	journal = "Phys. Rev. Lett.",
	volume = "88",
	pages = "191301",
	year = "2002"
}

@article{Merritt2004,
	author = "Merritt, David",
	title = "{Evolution of the Dark Matter Distribution at the Galactic Center}",
	doi = "https://doi.org/10.1103/PhysRevLett.92.201304",
	journal = "Phys. Rev. Lett.",
	volume = "92",
	pages = "201304",
	year = "2004"
}

@article{Bertone2005b,
	author = "Bertone, Gianfranco and Merritt, David",
	title = "{Time-dependent models for dark matter at the galactic center}",
	doi = "https://doi.org/10.1103/PhysRevD.72.103502",
	journal = "Phys. Rev. D",
	volume = "72",
	pages = "103502",
	year = "2005"
}

@article{Sadeghian2013,
	author = "Sadeghian, Laleh and Ferrer, Francesc and Will, Clifford M.",
	title = "{Dark-matter distributions around massive black holes: A general relativistic analysis}",
	doi = "https://doi.org/10.1103/PhysRevD.88.063522",
	journal = "Phys. Rev. D",
	volume = "88",
	number = "6",
	pages = "063522",
	year = "2013"
}

@article{Ferrer2017,
	author = "Ferrer, Francesc and Medeiros da Rosa, Alexandre and Will, Clifford M.",
	title = "{Dark matter spikes in the vicinity of Kerr black holes}",
	doi = "https://doi.org/10.1103/PhysRevD.96.083014",
	journal = "Phys. Rev. D",
	volume = "96",
	number = "8",
	pages = "083014",
	year = "2017"
}

@article{Shapiro2016,
	author = "Shapiro, Stuart L. and Shelton, Jessie",
	title = "{Weak annihilation cusp inside the dark matter spike about a black hole}",
	doi = "https://doi.org/10.1103/PhysRevD.93.123510",
	journal = "Phys. Rev. D",
	volume = "93",
	number = "12",
	pages = "123510",
	year = "2016"
}

@article{Xiong2026,
	author = "Xiong, W. and Li, P.-C.",
	title = "{Black hole spacetimes with dark matter spikes: Energy-momentum tensor and backreaction effects}",
	doi = "https://doi.org/10.1103/PhysRevD.113.124045",
	journal = "Phys. Rev. D",
	volume = "113",
	number = "12",
	pages = "124045",
	year = "2026"
}

@article{Einstein1939,
	author = "Einstein, Albert",
	title = "{On a Stationary System With Spherical Symmetry Consisting of Many Gravitating Masses}",
	doi="https://doi.org/10.2307/1968902",
	journal = "Ann. Math.",
	volume = "40",
	pages = "922--936",
	year = "1939"
}

@article{Bohmer2007,
	author = "B{\"o}hmer, Christian G. and Harko, Tiberiu",
	title = "{On Einstein clusters as galactic dark matter haloes}",
	doi="https://doi.org/10.1111/j.1365-2966.2007.11977.x",
	journal = "Mon. Not. R. Astron. Soc.",
	volume = "379",
	pages = "393--398",
	year = "2007"
}

@article{Jusufi2023,
	author = "Jusufi, Kimet",
	title = "{Black holes surrounded by Einstein clusters as models of dark matter fluid}",
	doi="https://doi.org/10.1140/epjc/s10052-023-11264-w",
	journal = "Eur. Phys. J. C",
	volume = "83",
	pages = "103",
	year = "2023"
}

@article{Acharyya2024,
	author = "Acharyya, R. and Banerjee, P. and Kar, S.",
	title = "{Modelling Einstein cluster using Einasto profile}",
	doi = "https://doi.org/10.1088/1475-7516/2024/04/070",
	journal = "J. Cosmol. Astropart. Phys.",
	volume = "2024",
	number = "04",
	pages = "070",
	year = "2024"
}

@article{Shen2024,
	author = "Shen, Z. and Wang, A. and Gong, Y. and Yin, S.",
	title = "{Analytical models of supermassive black holes in galaxies surrounded by dark matter halos}",
	doi="https://doi.org/10.1016/j.physletb.2024.138797",
	journal = "Phys. Lett. B",
	volume = "855",
	pages = "138797",
	year = "2024"
}

@article{Shen2025,
	author = "Shen, Z. and Wang, A. and Yin, S.",
	title = "{Inner radius and energy conditions of dark matter halos surrounding Schwarzschild black holes}",
	doi="https://doi.org/10.1016/j.physletb.2025.139300",
	journal = "Phys. Lett. B",
	volume = "862",
	pages = "139300",
	year = "2025"
}

@article{Maeda2025,
	author = "Maeda, Kei-ichi and Cardoso, Vitor and Wang, Anzhong",
	title = "{Einstein cluster as central spiky distribution of galactic dark matter}",
	doi = "https://doi.org/10.1103/PhysRevD.111.044060",
	journal = "Phys. Rev. D",
	volume = "111",
	pages = "044060",
	year = "2025"
}

@article{Geralico2012,
	author = "Geralico, A. and Pompi, F. and Ruffini, R.",
	title = "{On Einstein clusters}",
	doi="https://doi.org/10.1142/S2010194512006356",
	journal = "Int. J. Mod. Phys. Conf. Ser.",
	volume = "12",
	pages = "146--173",
	year = "2012"
}

@article{Lacroix2018,
	author = "Lacroix, Thomas",
	title = "{Dynamical constraints on a dark matter spike at the Galactic centre from stellar orbits}",
	doi="https://doi.org/10.1051/0004-6361/201832652",
	journal = "Astron. Astrophys.",
	volume = "619",
	pages = "A46",
	year = "2018"
}

@article{Shen2024b,
	author = "Shen, Z.-Q. and Yuan, G.-W. and Jiang, C.-Z. and Tsai, Y.-L. S. and Yuan, Q. and Fan, Y.-Z.",
	title = "{Exploring dark matter spike distribution around the Galactic centre with stellar orbits}",
	doi="https://doi.org/10.1093/mnras/stad3282",
	journal = "Mon. Not. R. Astron. Soc.",
	volume = "527",
	pages = "3196--3207",
	year = "2024"
}

@article{GRAVITY2024,
	author = "{GRAVITY Collaboration} and Abd El Dayem, K. and others",
	title = "{Improving constraints on the extended mass distribution in the Galactic center with stellar orbits}",
	doi="https://doi.org/10.1051/0004-6361/202452274",
	journal = "Astron. Astrophys.",
	volume = "692",
	pages = "A242",
	year = "2024"
}

@article{Nampalliwar2021,
	author = "Nampalliwar, S. and Kumar, S. and Jusufi, K. and Wu, Q. and Jamil, M. and Salucci, P.",
	title = "{Modeling the Sgr A* Black Hole Immersed in a Dark Matter Spike}",
	doi="https://doi.org/10.3847/1538-4357/ac05cc",
	journal = "Astrophys. J.",
	volume = "916",
	pages = "116",
	year = "2021"
}

@article{Daghigh2024,
	author = "Daghigh, R. G. and Kunstatter, G.",
	title = "{Effect of dark matter on galactic black hole ringdown waveforms and shadows}",
	doi = "https://doi.org/10.1103/PhysRevD.109.083004",
	journal = "Phys. Rev. D",
	volume = "109",
	pages = "083004",
	year = "2024"
}

@article{Daghigh2022,
	author = "Daghigh, R. G. and Kunstatter, G.",
	title = "{Spacetime Metrics and Ringdown Waveforms for Galactic Black Holes Surrounded by a Dark Matter Spike}",
	doi="https://doi.org/10.3847/1538-4357/ac940b",
	journal = "Astrophys. J.",
	volume = "940",
	pages = "33",
	year = "2022"
}

@article{Zhao2023,
	author = "Zhao, Y. and Sun, B. and Lin, K. and Cao, Z.",
	title = "{Axial gravitational ringing of a spherically symmetric black hole surrounded by dark matter spike}",
	doi = "https://doi.org/10.1103/PhysRevD.108.024070",
	journal = "Phys. Rev. D",
	volume = "108",
	pages = "024070",
	year = "2023"
}

@article{Speeney2024,
	author = "Speeney, N. and Berti, E. and Cardoso, V. and Maselli, A.",
	title = "{Black holes surrounded by generic matter distributions: Polar perturbations and energy flux}",
	doi = "https://doi.org/10.1103/PhysRevD.109.084068",
	journal = "Phys. Rev. D",
	volume = "109",
	pages = "084068",
	year = "2024"
}

@article{Zhao2024,
	author = "Zhao, Y. and Sun, B. and Cao, Z. and Lin, K. and Qian, W.-L.",
	title = "{Influence of dark matter equation of state on the axial gravitational ringing of supermassive black holes}",
	doi = "https://doi.org/10.1103/PhysRevD.109.044031",
	journal = "Phys. Rev. D",
	volume = "109",
	pages = "044031",
	year = "2024"
}

@article{Eda2013,
	author = "Eda, K. and Itoh, Y. and Kuroyanagi, S. and Silk, J.",
	title = "{New Probe of Dark-Matter Properties: Gravitational Waves from an Intermediate-Mass Black Hole Embedded in a Dark-Matter Minispike}",
	doi = "https://doi.org/10.1103/PhysRevLett.110.221101",
	journal = "Phys. Rev. Lett.",
	volume = "110",
	pages = "221101",
	year = "2013"
}

@article{Eda2015,
	author = "Eda, K. and Itoh, Y. and Kuroyanagi, S. and Silk, J.",
	title = "{Gravitational waves as a probe of dark matter minispikes}",
	doi = "https://doi.org/10.1103/PhysRevD.91.044045",
	journal = "Phys. Rev. D",
	volume = "91",
	pages = "044045",
	year = "2015"
}

@article{Kavanagh2020,
	author = "Kavanagh, B. J. and Nichols, D. A. and Bertone, G. and Gaggero, D.",
	title = "{Detecting dark matter around black holes with gravitational waves: Effects of dark-matter dynamics on the gravitational waveform}",
	doi = "https://doi.org/10.1103/PhysRevD.102.083006",
	journal = "Phys. Rev. D",
	volume = "102",
	pages = "083006",
	year = "2020"
}

@article{Hassanabadi2026,
	author = "Hassanabadi, H. and Zare, Soroush",
	title = "{Black holes in depleted Dehnen dark matter halos with physical inner edges and strong-field observables}",
	eprint = "2609.23750",
	archivePrefix = "arXiv",
	primaryClass = "gr-qc",
	journal = "arXiv:2609.23750",
	year = "2026"
}

@article{Bowers1974,
	author = "Bowers, Richard L. and Liang, E. P. T.",
	title = "{Anisotropic spheres in general relativity}",
	doi="10.1086/152760",
	journal = "Astrophys. J.",
	volume = "188",
	pages = "657--665",
	year = "1974"
}

@article{Israel1966,
	author = "Israel, Werner",
	title = "{Singular hypersurfaces and thin shells in general relativity}",
	doi = "https://doi.org/10.1007/BF02710419",
	journal = "Nuovo Cim. B",
	volume = "44",
	pages = "1--14",
	year = "1966"
}

@article{Keeton2005,
	author = "Keeton, Charles R. and Petters, Arlie O.",
	title = "{Formalism for testing theories of gravity using lensing by compact objects: Static, spherically symmetric case}",
	doi = "https://doi.org/10.1103/PhysRevD.72.104006",
	journal = "Phys. Rev. D",
	volume = "72",
	pages = "104006",
	year = "2005"
}

@article{Bozza2002,
	author = "Bozza, Valerio",
	title = "{Gravitational lensing in the strong field limit}",
	doi = "https://doi.org/10.1103/PhysRevD.66.103001",
	journal = "Phys. Rev. D",
	volume = "66",
	pages = "103001",
	year = "2002"
}

@article{Cardoso2009,
	author = "Cardoso, Vitor and Miranda, Alex S. and Berti, Emanuele and Witek, Helvi and Zanchin, Vilson T.",
	title = "{Geodesic stability, Lyapunov exponents, and quasinormal modes}",
	doi = "https://doi.org/10.1103/PhysRevD.79.064016",
	journal = "Phys. Rev. D",
	volume = "79",
	pages = "064016",
	year = "2009"
}

@article{Leung1999,
	author = "Leung, P. T. and Liu, Y. T. and Suen, W.-M. and Tam, C. Y. and Young, K.",
	title = "{Perturbative approach to the quasinormal modes of dirty black holes}",
	doi = "https://doi.org/10.1103/PhysRevD.59.044034",
	journal = "Phys. Rev. D",
	volume = "59",
	pages = "044034",
	year = "1999"
}

@article{IyerWill1987,
	author = "Iyer, Sai and Will, Clifford M.",
	title = "{Black-hole normal modes: A WKB approach. I. Foundations and application of a higher-order WKB analysis of potential-barrier scattering}",
	doi = "https://doi.org/10.1103/PhysRevD.35.3621",
	journal = "Phys. Rev. D",
	volume = "35",
	pages = "3621--3631",
	year = "1987"
}

@article{Schutz1985,
	author = "Schutz, Bernard F. and Will, Clifford M.",
	title = "{Black hole normal modes: A semianalytic approach}",
	doi = "https://doi.org/10.1086/184453",
	journal = "Astrophys. J.",
	volume = "291",
	pages = "L33--L36",
	year = "1985"
}

@article{Iyer1987,
	author = "Iyer, Sai",
	title = "{Black-hole normal modes: A WKB approach. II. Schwarzschild black holes}",
	doi = "https://doi.org/10.1103/PhysRevD.35.3632",
	journal = "Phys. Rev. D",
	volume = "35",
	pages = "3632--3636",
	year = "1987"
}

@article{Konoplya2003,
	author = "Konoplya, Roman A.",
	title = "{Quasinormal behavior of the D-dimensional Schwarzschild black hole and the higher order WKB approach}",
	doi = "https://doi.org/10.1103/PhysRevD.68.024018",
	journal = "Phys. Rev. D",
	volume = "68",
	pages = "024018",
	year = "2003"
}

@article{Konoplya2019,
	author = "Konoplya, R. A. and Zhidenko, A. and Zinhailo, A. F.",
	title = "{Higher order WKB formula for quasinormal modes and greybody factors: Recipes for quick and accurate calculations}",
	doi = "10.1088/1361-6382/ab2e25",
	journal = "Class. Quantum Grav.",
	volume = "36",
	pages = "155002",
	year = "2019"
}

@article{Harmark2010,
	author = "Harmark, Troels and Nat{\'a}rio, Jos{\'e} and Schiappa, Ricardo",
	title = "{Greybody factors for d-dimensional black holes}",
	doi="10.4310/ATMP.2010.v14.n3.a1",
	journal = "Adv. Theor. Math. Phys.",
	volume = "14",
	pages = "727--794",
	year = "2010"
}

\end{document}